\documentclass[aps,preprint,groupedaddress,nofootinbib,longbibliography,floatfix]{revtex4-1}
\usepackage{amsmath}
\usepackage{amsfonts}
\usepackage{graphicx}
\usepackage{color}

\def\dalemb#1#2{{\vbox{\hrule height.#2pt
      \hbox{\vrule width.#2pt height#1pt \kern#1pt \vrule width.#2pt}
      \hrule height.#2pt}}}

\def\ba{\begin{eqnarray}}
\def\ea{\end{eqnarray}}
\def\be{\begin{equation}}
\def\ee{\end{equation}}

\def\gtorder{\mathrel{\raise.3ex\hbox{$>$}\mkern-14mu
           \lower0.6ex\hbox{$\sim$}}}
\def\ltorder{\mathrel{\raise.3ex\hbox{$<$}\mkern-14mu
           \lower0.6ex\hbox{$\sim$}}}

\def\C{\mathbf{C}}
\def\N{\mathbf{N}}

\def\f{\mathbf{f}}

\newcommand{\half}{\frac{1}{2}}

\def\betab{\boldsymbol{\beta }}

\begin{document}

\title{Reconstructing the primordial power
spectrum from the CMB}

\author{Christopher Gauthier and Martin Bucher}

\affiliation{Laboratoire APC,
Universit\'e Paris Diderot,
B\^atiment Condorcet,\\
10 rue Alice Domon et L\'eonie Duquet,
75205 Paris Cedex 13, France}

\begin{abstract}
We propose a straightforward and model independent methodology
for characterizing the sensitivity of CMB and other experiments
to wiggles, irregularities, and features in the primordial power spectrum.
Assuming that the primordial cosmological perturbations are adiabatic, we
present a function space generalization of the usual Fisher matrix
formalism, applied to a CMB 
experiment resembling Planck with and without ancillary data. This
work is closely related to other work on recovering the inflationary
potential and exploring specific models of non-minimal, or perhaps 
{\it baroque}, primordial power spectra. The approach adopted here, 
however, most directly expresses what the data is really telling us.
We explore in detail the structure of the available information and
quantify exactly what features can be reconstructed and at what statistical
significance.
\end{abstract}

\date{11 September 2012}

\maketitle

\section{Introduction}

In the last decade, CMB experiments have created new opportunities 
for studying the early universe. With the data obtained 
by WMAP \cite{Komatsu:2010fb} and the promise of even more accurate CMB measurements 
by Planck \cite{bluebook} in addition to the several current and future ground 
and balloon based experiments, 
the data will become even more constraining and it will be interesting to see whether
the simple minimal model that seems to be able to explain all the present data survives
or whether a new layer of complexity is discovered.

According to our current best understanding of primordial cosmology,
the universe underwent 
a period of extremely rapid expansion known as inflation 
\cite{Guth:1980zm,Linde:1981mu,Linde:1983gd,Albrecht:1982wi}. This 
model was initially proposed to solve several problems with the standard 
big bang theory. In addition, inflation was found to make quantitative 
predictions concerning the primordial cosmological perturbations \cite{
Hawking:1982cz,
Bardeen:1983qw,
Guth:1982ec,
Mukhanov:1981xt}.
In the 
simplest of these models, inflation predicts that the primordial 
fluctuations in primordial power spectrum (PPS) should be Gaussian with a nearly scale invariant 
spectrum. 
The most common method of interpreting the CMB data assumes a modest 
number of cosmological parameters (for example, in the WMAP 7-year 
analysis the parameters adopted are $H_0,$ $\omega _b,$ $\omega _c,$ 
$\Omega _\Lambda $) and a parameterized form for the primordial power 
spectrum, for example $P(k)=A(k/k_0)^{n_s-1}$ or perhaps 
$P(k)=A(k/k_0)^{(n_s-1-\beta \ln(k/k_0))}$ where the parameter $\beta $ 
denotes the running of the spectral index. In this paper we adopt as a 
fiducial reference model the WMAP 7-year best fit \cite{Komatsu:2010fb}, 
for which the parameters are $h = 0.703$, $\omega _b=h^2\Omega 
_b=0.02227$, $\omega _c=h^2\Omega _c=0.1116$, $\tau =0.085$, $A=2.42 
\times 10^{-9}$, and $n_S=0.966$. Here the somewhat arbitrary pivot 
scale has been set to $k_0=2.0\times 10^{-3}~Mpc^{-1}$. Furthermore we 
assume that there are no tensor perturbations nor running in the scalar 
spectral index (i.e. $r=0$ and $\beta = 0.0$) and also that the 
universe is exactly flat. It is encouraging to learn that a simple model 
of this sort can provide an adequate and plausible explanation of the 
presently available data, and following this general approach, one can 
construct more complicated models and compare 
these based on relative likelihood and ideas about model comparison.

However the drawback of an analysis comparing parameterized models to 
the data and among themselves is that which aspects of the model actually 
enter into the likelihood remains hidden inside a black box. 
If for example we show plots of constraints on $n_s$ or the running
$\beta ,$ these plots do not show at what $k$ measurements have been
made.  The purpose 
of this paper is to enter into this black box and show numerically what 
exactly is being measured, in particular with respect to the primordial 
power spectrum. For simplicity we assume that the perturbations are 
adiabatic and Gaussian, although generalizing the treatment here to 
isocurvature perturbations is in principle straightforward.

Rather than developing a suite of fancy parametric 
models with lots of bells and whistles, we approach the problem more in 
the spirit of the method dubbed `exploratory data analysis' 
by John Tukey 
(see for example \cite{Tukey1977}) where the aim is to 
devise methods to make manifest patterns in the data and what is being 
constrained and where, the purpose being to uncover the unexpected. 
This approach stands in contrast to the complementary approach where 
precise models with well-defined prior information are set forth as 
hypotheses and optimal statistical tests to discriminate between them
are sought. In fact the use of 
priors in the paper is more formal and metaphorical, because for our 
purposes meaningful prior information is lacking. The {\it priors} serve 
more as regulators to suppress noise where there is seemingly no 
relevant information.

In the past a number of approaches have been proposed to search for 
features in the primordial power spectrum inferred from the CMB. One 
general class of approaches consists of parametric models, where $P(k)$ 
is modeled using a family of functions having a fixed and 
well-defined number of adjustable parameters. Bridle et al. 
\cite{Bridle:2003sa}, for example, analyzed the WMAP first-year results 
using a model with N knots equally spaced in $\ln (k)$ between a 
$k_{min}$ and $k_{max}$ where $\ln (P(k))$ is interpolated linearly 
between $N$ control points or knots. The number of knots admitted is a 
measure of the model complexity, and as is well-known, 
when comparing models it is necessary 
to compensate for the better fit due to fitting away 
noise when more degrees of freedom are present. This can be 
done using standard techniques (e.g., a simple $\chi ^2$ analysis, the 
Aikake information criterion (AIC) \cite{Aikake1974}, the Bayesian 
information criterion (BIC) \cite{Schwarz1978}\footnote{ 
\baselineskip=0pt Care must be taken when applying the BIC to actual 
data because of the asymptotic nature of its derivation. It is often 
asked what value to use for $n_{data}$ in the $\log (n_{data})$ term 
appearing in the BIC. Is it the number of $C_\ell $ or the total number 
of measurements? A careful analysis starting from a Bayesian prior 
reveals that while the BIC obtains the right scaling as 
$n_{data}\to \infty ,$ 
there is another prefactor that does not vary with $n_{data}$ 
that is ignored. Without knowing this prefactor, which would required a 
specific Bayesian prior, it is not possible to address this question. In 
reality, the factor should be $\ln (n_{data}/\bar n)$ where $\bar n$ 
depends on the overall relative likelihood of the models in the prior.
}
, etc). One disadvantage of this approach is that the placement 
of the bin boundaries, which is arbitrary, influences the conclusions.
The sensitivity to a sharp feature having a width comparable to 
the bin width depends on where the feature is situated relative to the 
bin boundaries.  Interesting variants have been proposed. Hamann et al.~have
proposed allowing the knot positions not to be fixed but rather parameters 
of the model \cite{Hamann:2009bz}. Although this approach introduces nonlinearity 
into the model, it allows localized sharp features to be 
modeled by placing control points only where they are needed. In this way models 
giving a good fit to a single sharp feature are significantly less 
penalized than in the approach of Bridle et al. and others 
\cite{Hannestad:2003zs,Bridle:2003sa}.

Another class of approaches may be described as penalized likelihood 
methods. In this class of methods the function space for $P(k)$ 
includes many more parameters than desired. There is either a very large 
or infinite number of degrees of freedom. Model complexity is limited by 
penalizing very rough or rapidly varying functions. The log likelihood 
is modified to include a roughness penalty for example as in the 
following form:
\ba
\ln ~{L}({\bf d}, P(k), \boldsymbol{\beta })
+ \lambda 
\int d(\ln [k])~
\left(
\frac{
\partial ^2 
(\ln P)
}{
\partial 
(\ln k)^2
}
\right) ^2 .
\label{regulator}
\ea
Here we may appeal to Bayesian doctrine by describing the second term as 
a prior probability over the function space of $P(k)$. An example of this approach
applied to the WMAP data can be found in \cite{Sealfon:2005em}. However in practice 
this term rarely corresponds to any actual prior belief. Rather it is 
chosen {\it a posteriori} to obtain a meaningful result and depends on the 
quality of the data. The form of the roughness penalty in 
eqn.~(\ref{regulator}) is somewhat arbitrary and other reasonable forms have 
been proposed. We could also have used the square of the first rather 
than the second derivative, although this choice would have the drawback of 
slightly favoring an overall flatter spectrum. 
Also the weighting as a function of $\ln (k)$ and what is assumed for $k$
smaller and larger than where there is data involves arbitrary or subjective
choices.  The challenge of choosing 
the regularization parameter $\lambda $ involves finding the right 
compromise between underfitting and overfitting the data. There exists 
an abundant literature on how to make an {\it optimal} choice for $\lambda $ 
rapidly surveyed below. However in our view it does not suffice 
to find the {\it optimal} $\lambda ,$ because this information does not 
tell us which features of the reconstruction are to be 
believed and which are to be regarded as most likely explained as 
statistical fluctuations. The approach adopted in the paper is to study 
and characterize precisely the properties of the reconstruction for a 
variety of reasonable values for the smoothing parameter.

Tocchini-Valentini 
et al.~\cite{TocchiniValentini:2004ht,TocchiniValentini:2005ja}
adopted a penalized likelihood similar to
that used in this paper, which was applied to the WMAP data finding
some evidence for statistically significant features. In this paper
we account for uncertainties in the cosmological parameters rather than
fixing them and adopt a frequentist approach to the error analysis because
we do not believe that the prior implicit in the smoothing penalty deserves
to be interpreted as an actual prior belief or probability measure
on the function space of possible power spectra. 

Many reconstruction procedures can in some cases yield reconstructed
primordial power spectra that take negative values in some places,
which is clearly unphysical. This can occur when a quadratic approximation
to the likelihood is made or when there is a superimposed noise component.
Hamann et al.~\cite{Hamann:2009bz} 
(see also \cite{Shafieloo:2007tk}) carried out a modified Lucy-Richardson
deconvolution of the primordial power spectrum from CMB data.
Lucy \cite{1974AJ.....79..745L} and 
Richardson \cite{RICHARDSON:72}
introduced an algorithm widely used for
image deconvolution having the property that the reconstruction
is always positive. The original Lucy-Richardson algorithm was
derived from a likelihood assuming a Poisson noise for the number
counts. If there are some counts in a cell, an infinitely improbable
likelihood results as the predicted average count tends to zero,
quite unlike in a Gaussian, where a negative average count would not be excluded
by the form of the likelihood.
Unfortunately the CMB likelihood does not fit into this mold. 
Modifications to L-R algorithm have been proposed to take into
account the properties of the CMB likelihood. However the modified
L-R algorithm cannot be derived rigorously starting from a CMB
likelihood.  Hamann et al.~overcome this difficulty by carrying out a 
frequentist analysis of the reconstructed power spectra, treating
the reconstruction as a black box and characterizing its properties
through Monte Carlo experiments.

Hu \cite{Hu:2003vp}
and Leach \cite{Leach:2005av}
 explore an approach based on principal component 
analysis. A principal component analysis ranks directions of deviation
with respect to a fiducial power spectrum using the eigenvalues
and eigenvectors of the Fisher information matrix. The decomposition
into eigenmodes, however, relies on a notion of length (or metric)
for this space of linearized deviations, and thus entails some 
arbitrariness.\footnote{A quadratic form is a linear mapping
from a linear space into its dual and not into itself. Therefore
a metric tensor is required to define a natural equivalence between
the linear space and its dual.}
Peiris et al.~\cite{Sealfon:2005em,Verde:2008zza,Peiris:2009wp} 
explore a penalized likelihood using `cross-validation' to select
an optimal degree of smoothing. 
A variety of other approaches have been proposed
including wavelets
(Mukherjee et al.~\cite{Mukherjee:2003yx,Mukherjee:2003ag,Mukherjee:2005dc})
and `cosmic inversion' (Kogo et al.~\cite{Kogo:2003yb},
Nagata et al.~\cite{Nagata:2008tk,Nagata:2008zj} and 
Ichiki et al.~\cite{Ichiki:2009zz}).

Vazquez et al.~\cite{Vazquez:2012ux} present an interesting analysis 
based on Bayesian model comparison. A prior distribution is constructed 
for a linear interpolation of $\ln (P)$ as a function of $\ln (k)$
using a variable number of nodes whose positions are
free, yielding a posterior distribution among the models. Using data
from WMAP combined with ACT and SDSS LRG data, the authors conclude
that an $n_S=1$ Harrison-Zeldovich-Peebles spectrum is `strongly'
excluded within a Bayesian framework, using the terminology
of Jeffreys. 

This paper systematically studies and characterizes the trade-offs 
between bias and variance (i.e., noise) in the choice of smoothing 
scheme. Rather than espousing a particular philosophy leading to a 
unique solution, we show in a quantitative way the results for a range 
of choices. Which is best depends on what type of features one is trying 
to reconstruct.

The organization of the paper is as follows. Section 2 presents a detailed
description of the constraints on the primordial power spectrum
assuming a Planck-like experiment based on the Fisher information kernel.
The information loss due to uncertainty in the determination of the 
other cosmological parameters is described.
Section 3 describes the power spectrum construction first for the simplified
case where the other cosmological parameters are fixed and then including
their uncertainties. 
Section 4 characterizes the noise of the reconstruction and with what
statistical significance various model features can be reconstructed.
Finally Section 5 presents some concluding remarks.

\clearpage
\section{Fisher Information Kernel}

\begin{figure}[htpb]
\begin{center}
\includegraphics[scale=.5]{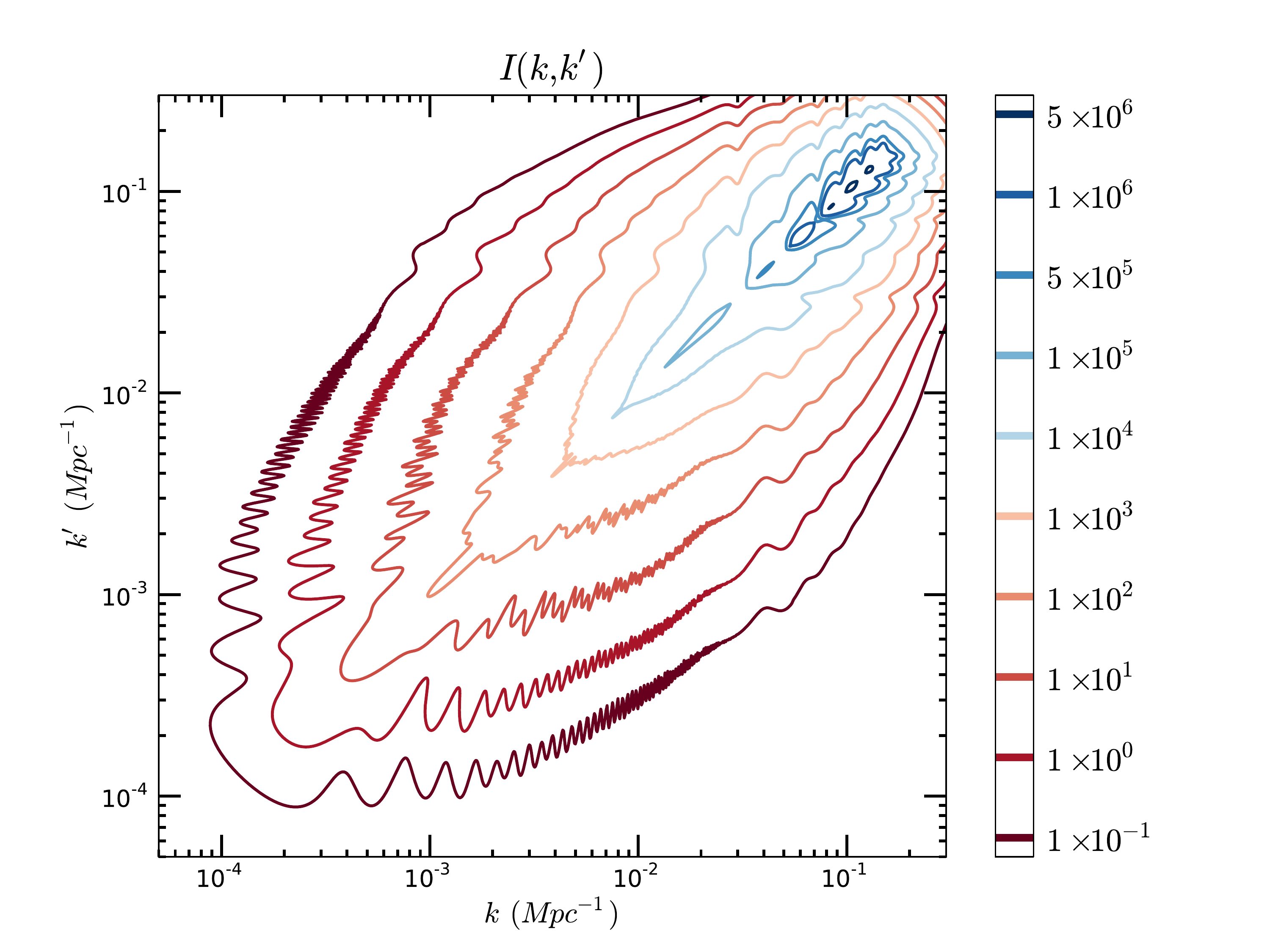}
\caption{\baselineskip=0pt
{\bf Fisher information density relative to a reference
primordial power spectrum.}
The information density defined in eqn.~(\ref{FI:def}) is shown.
}
\label{contour_Kern}
\end{center}
\end{figure}

\begin{figure}[!htb]
\begin{center}
\includegraphics[width=14cm]{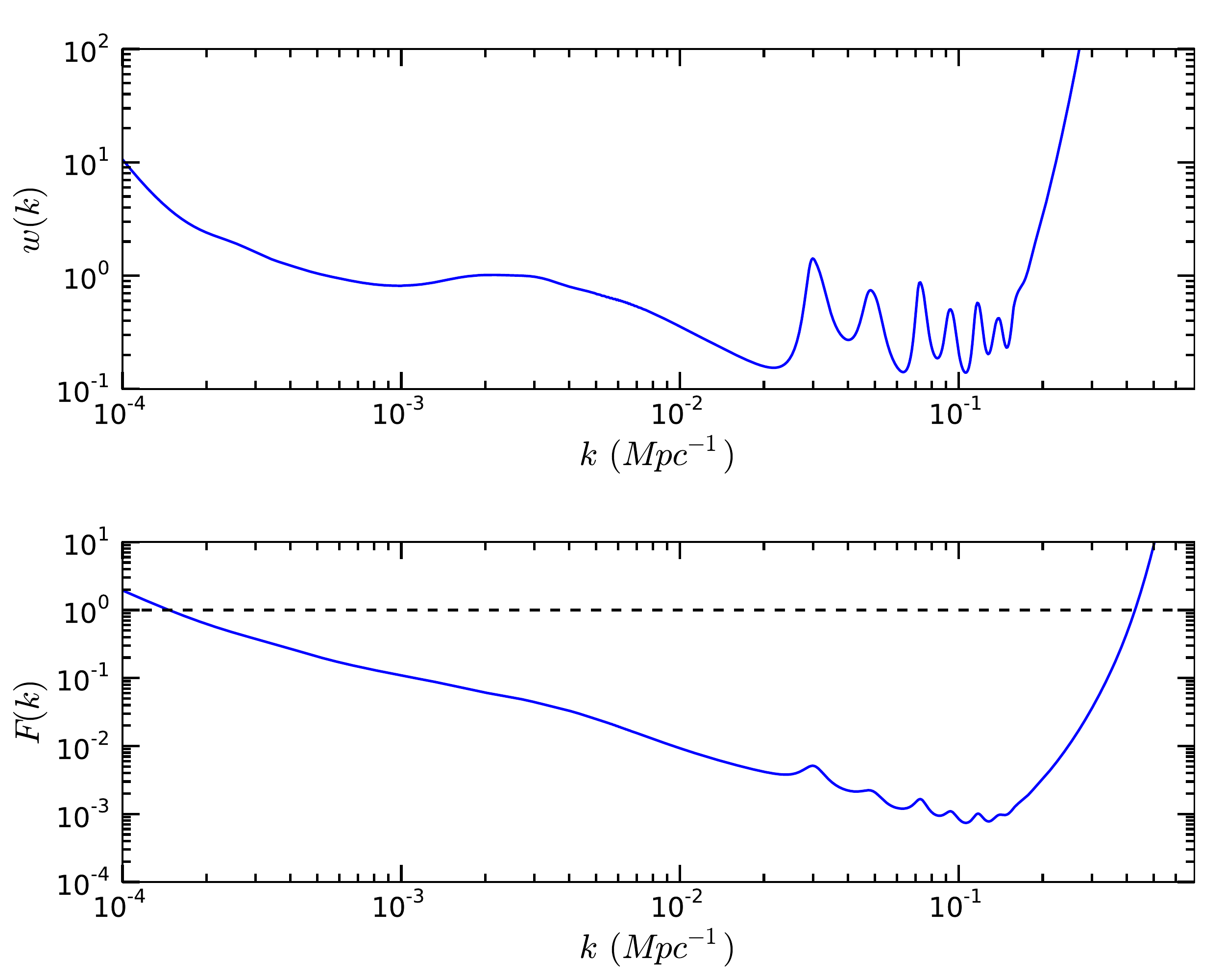}
\includegraphics[scale=.45]{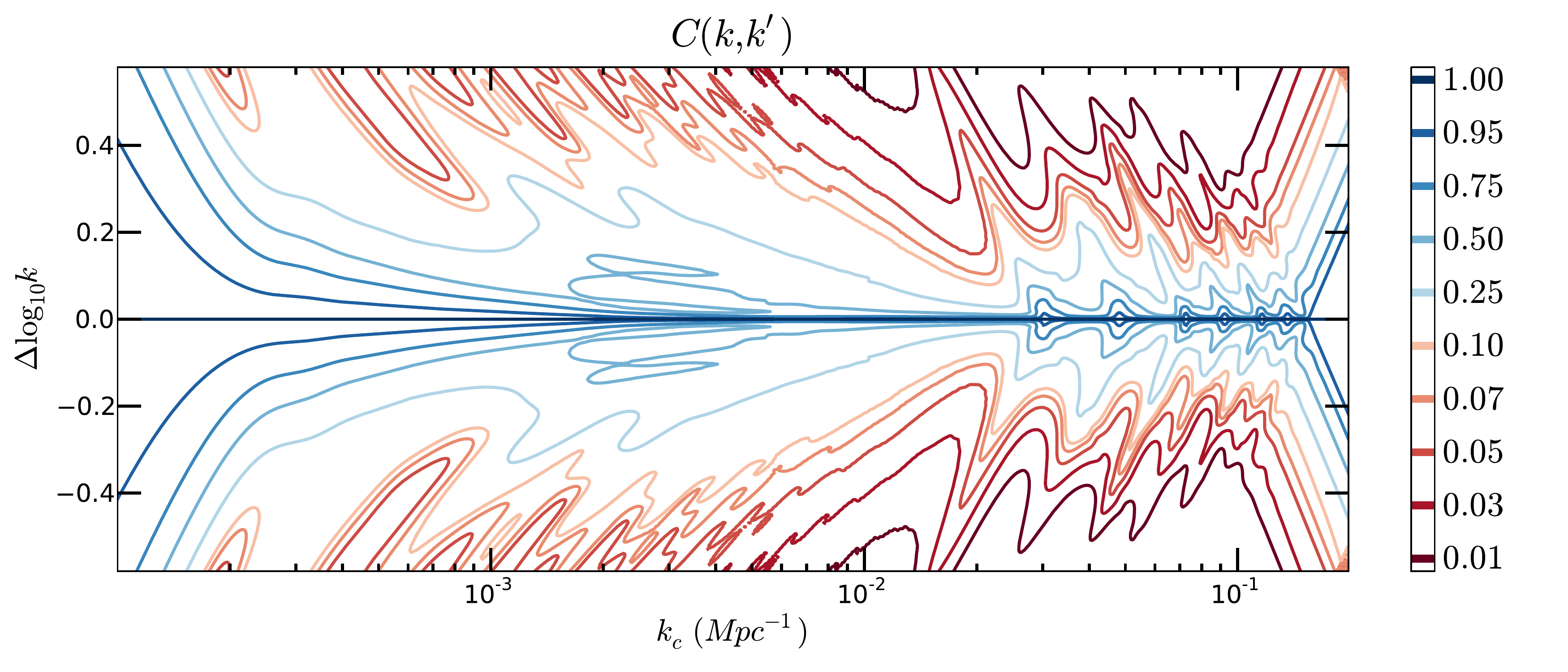}
\caption{\baselineskip=0pt
{\bf Properties of the Fisher information density kernel.}
The top plot shows 
the width as defined in eqn.~(\ref{WidthDef}). 
The middle plot shows
$F(k)$ as defined in eqn.~(\ref{II:info}),
which provides an estimate of how accurately fluctuations at the 
scale $k$ are being constrained.
The bottom plot shows the structure of
the correlation matrix (i.e., the kernel $C(k,k')=I(k,k')/[I(k,k)I(k'k')]^{1/2}).$
Here horizontal coordinate is $k_{hor}=\sqrt{kk'}$ and the vertical coordinate
is $\ln (k'/k).$ On the far left and the far right the elements far away
from the diagonal have a correlation of almost one, indicating that only a single
quantity is being measured at very small and very large $k.$
}
\label{Corr:Fig}
\end{center}
\end{figure}

The full-sky temperature power spectrum $C_{\ell}$ depends linearly on the 
primordial power spectrum and is defined as the following 
linear integral transform
\begin{gather}
C_{\ell}^{TT}
=
\int_{-\infty}^{\infty}
d \ln k\,
P(k)
[\Delta _{\ell}^{(T)}(k)]^{2}
\end{gather}
where $\Delta _{\ell}^{(T)}(k)$ is an adiabatic temperature transfer function 
as computed using standard Boltzmann solver such as CAMB \cite{CAMB}.
The polarization and 
cross spectra,  
$C_{\ell}^{EE}$ and 
$C_{\ell}^{TE},$ 
are similarly defined. We approximate the 
likelihood by assuming full sky coverage and uncorrelated white noise
\begin{gather}
-\ln L
=
\frac{f_{sky}}{2}
\sum_{\ell=2}^{\ell_{max}}
(2 \ell + 1)
\left[
\textrm{tr}\left(
\frac{\mathbf{C}_\ell ^{(obs)}}{{\C}_{\ell}^{(th)}+ \N_{\ell}}
\right)
-
\ln\det
\left(
\frac{\mathbf{C}_\ell ^{(obs)}}{{\C}_{\ell}^{(th)} + \N_{\ell}}
\right)
-3
\right]
\label{cmblikelihood}
\end{gather}
where $\C_{\ell}^{(obs)}=\{ C_\ell ^{(obs),AB}\} _{A,B=T,E}$ is the measured
power spectrum matrix,
$\C _\ell ^{(th)}$ the theoretical (predicted) power spectrum matrix, 
and $\N _\ell $ is the detector noise. In this section and elsewhere in the 
paper unless otherwise specified, we assume a fictitious experiment 
somewhat similar to Planck based on the specifications
published in the bluebook \cite{bluebook}
rather than on an updated understanding of the mission and the instrument
in-flight performance. 

In this paper we 
are concerned with the effect that small scale variations in the 
primordial power spectrum have 
on the likelihood. We define the Fisher information kernel $I$ as
\begin{gather}
-2 \ln L 
= 
\int d\ln k 
\int d\ln k'
\left[
P(k)
-
P_{ML}(k)
\right]
I(k , k')
\left[
P(k')
-
P_{ML}(k')
\right]
\label{LikeEqq}
\end{gather}
where an irrelevant constant term has been omitted.
Using this definition we can express the kernel as a functional derivative of the log likelihood
\begin{gather}
I(\ln k, \ln k')
=
\left.
\frac{\delta^{2} (- \ln L)}{\delta P(\ln k) \delta P(\ln k')}
\right|_{P = P_{ML}}
\label{Idef}
\end{gather}
where $P_{ML}$ is the primordial power spectrum at maximum likelihood. 

Since we already have a rough idea of what the primordial power spectrum
looks like on the scales probed by the CMB, it makes sense to describe
the power spectra under consideration in terms of fractional variations
with respect to a fiducial power law spectrum (for which the parameters
are taken from the WMAP 7-year best fit model). We set
$P_0(k)=A(k/k_0)^{n_S-1}$ where $A=2.42\times 10^{-9},$
$k_0=2.0\times 10^{-3} Mpc^{-1},$ and $n_S=0.966,$ and define 
$P(k)=P_0(k) [1+ f(k)].$
Let us for the moment fix the other cosmological parameters to their
WMAP 7-year best fit values. We may express the expectation 
value of the relative log likelihood as the following quadratic form
about the best-fit value as follows:
\ba
\Delta [-\ln L]=
\half 
\int d(\ln k_1)
\int d(\ln k_2)
[ f(k_1) - f_{ML}(k_{1})
]
\hat I(k_1, k_2)
[
f(k_2)
-
f_{ML}(k_{2})
]
\label{FI:def}
\ea
where $\hat{I}(k, k') = 
P_{0}(k')
I(k, k') 
P_{0}(k) 
$. 
In our definitions we have tried
to remove dimensions from quantities
wherever possible. 
Fig.~\ref{contour_Kern} shows the Fisher information density
$\hat I(k_1, k_2).$ 
The integral over both arguments is interpreted 
as the $\chi ^2$ for detecting a CMB signal over a null
hypothesis of no CMB signal at all, and as expected this
quantity is approximately $10^6$ or $\ell _{max}^2$ where 
$\ell _{max}$ is approximately the multipole number
where the signal-to-noise ratio $(S/N)$ has dropped to one. 
Were it not for correlations in the Fisher information 
density, we would be able to measure about a million independent
quantities characterizing the shape of the power spectrum
at a marginal statistical significance---that is, at about
$1\sigma .$ This would indeed be the case if the Fisher
information density were more strongly peaked along the diagonal.
The width of the peaked region along the
diagonal provides an approximate idea of how coarsely
one has to bin for correlations between adjacent bins to
be negligible. If the peaked region along the diagonal is
very wide, we measure a small number of quantities (much
smaller than $\chi ^2_{total}$) albeit at very high precision.
The most extreme case would be a Fisher matrix where all
the entries are equal, or more generally of the form
$\hat I(\kappa _1,\kappa _2)= g(\kappa _1) g(\kappa _2)$
where we define $\kappa =\ln (k).$ In this case the Fisher
information operator would be of rank one and precisely
one quantity can be being measured. Fig.~\ref{Corr:Fig}
shows the diagonal of the Fisher information kernel
and the correlation matrix obtained from the 
Fisher information kernel---that is $C(\kappa , \kappa ')
=\hat I(\kappa , \kappa ')/
[\hat I(\kappa , \kappa ) \hat I(\kappa ', \kappa ')]^{1/2}.$
The fact that at very low $k$ and very high $k$ the correlation
becomes very nearly one even quite far from the diagonal
indicates that in each of these two region there is only one
relevant quantity that is being measured constituting a sort of 
weighted average of the power spectrum at each end.
\begin{figure}[!htbp]
\begin{center}
\includegraphics[width=15cm]{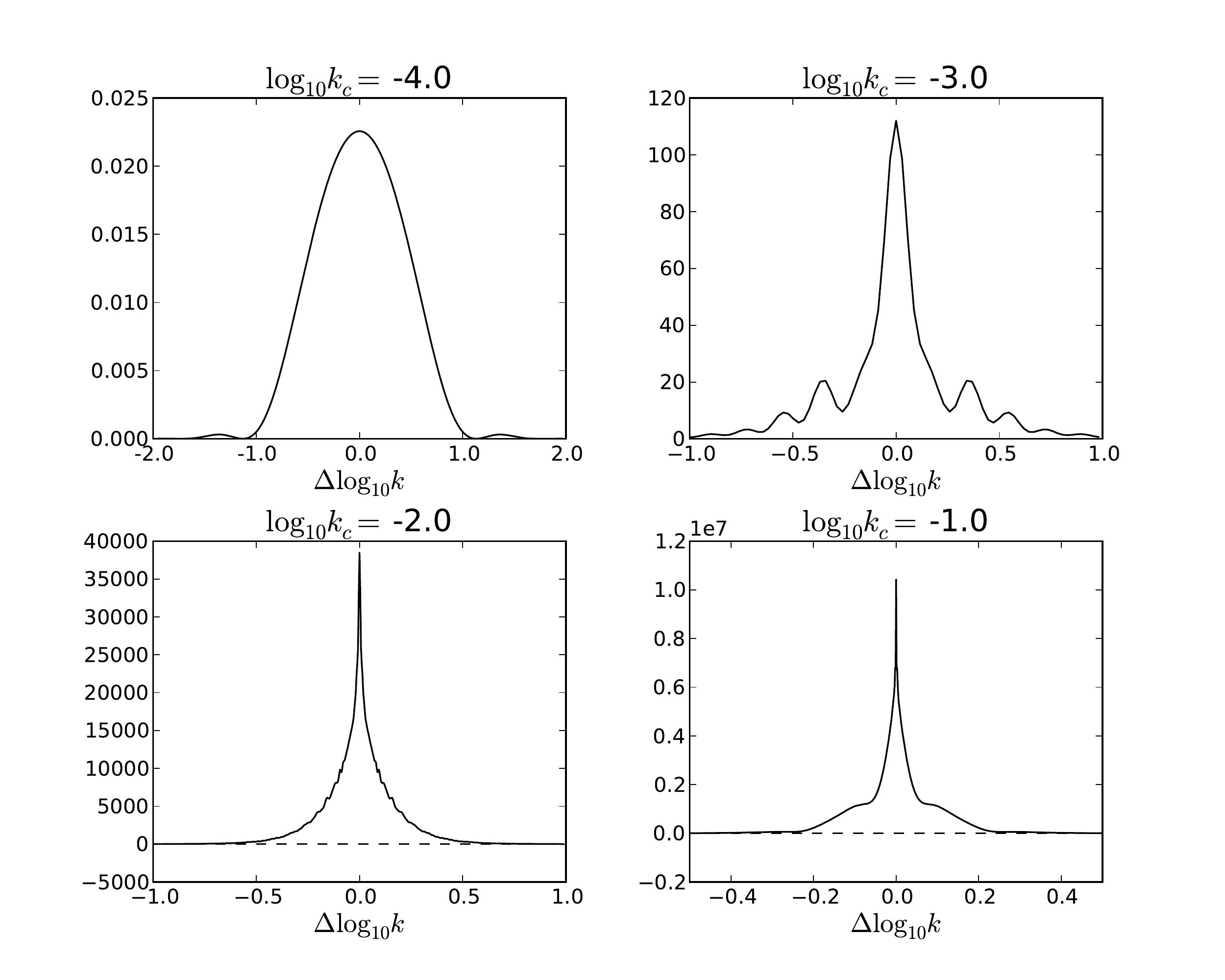}
\caption{
\baselineskip=0pt
{\bf Fisher density kernel 
transverse profile.} We plot the Fisher density kernel
in the direction orthogonal to the diagonal
(plots labeled according to wavenumber $k_{c}$ of diagonal crossing).
(i.e., $\hat I(k_c+\Delta k,k_c-\Delta k)$}
\label{transverse_xsec}
\end{center}
\end{figure}
It is convenient to define the local width of the diagonal in the following
way
\ba 
w(\ln k)=\frac{\int d(\ln k')~\hat I(\ln k, \ln k') }{ \hat I(\ln k, \ln k) }.
\label{WidthDef}
\ea
This quantity is plotted in Fig.~\ref{Corr:Fig} in the middle panel.
For $\Delta (\ln k)\gtorder w,$  the error in the binned power spectrum may
be approximated as 
\ba 
\sigma 
\left[
\frac{\delta P}{P}_{k\in [k_1,k_2]}
\right]
\approx 
\left[
\hat I\left( \half \ln k_1k_2, \half \ln k_1k_2\right)
w\left( \half \ln k_1k_2\right) \ln \left( \frac {k_2}{k_1}\right) 
\right]^{-1/2}.
\ea
Fig.~\ref{Corr:Fig}
shows (on a logarithmic scale)
\ba 
F(\ln k)=\left[ \hat I(\ln k, \ln k)
w(\ln k) \right] ^{-1/2}.
\label{II:info}
\ea
When $F\gtorder 1$ we have essentially no useful information regarding
the power spectrum given current expectations. 

\begin{figure}[!htbp]
\begin{center}
\includegraphics[scale=0.45]{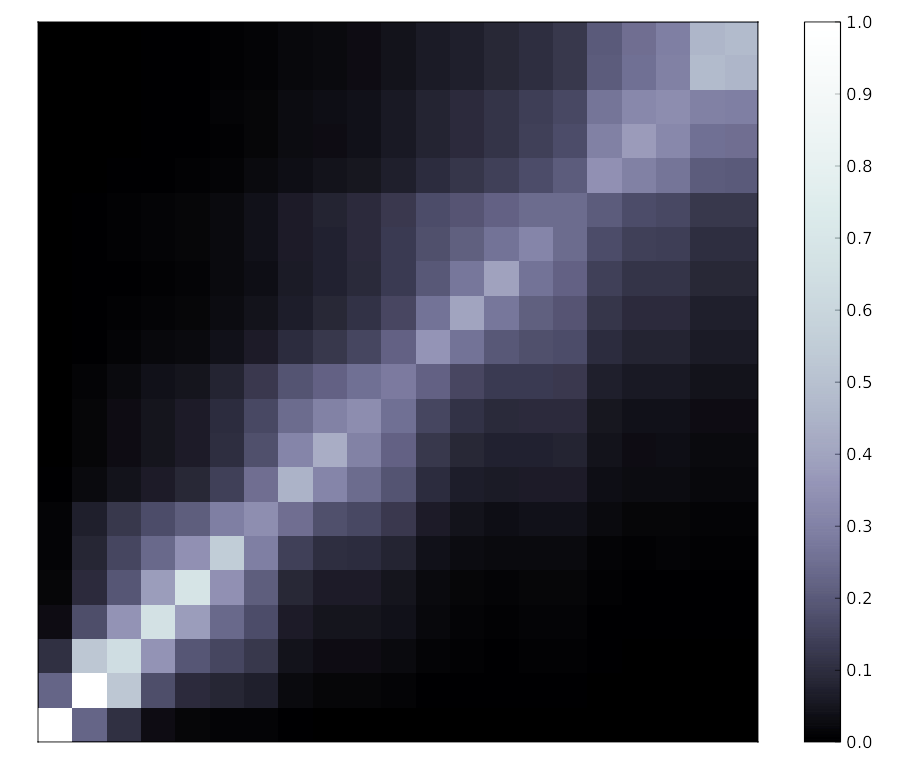}
\includegraphics[scale=0.3]{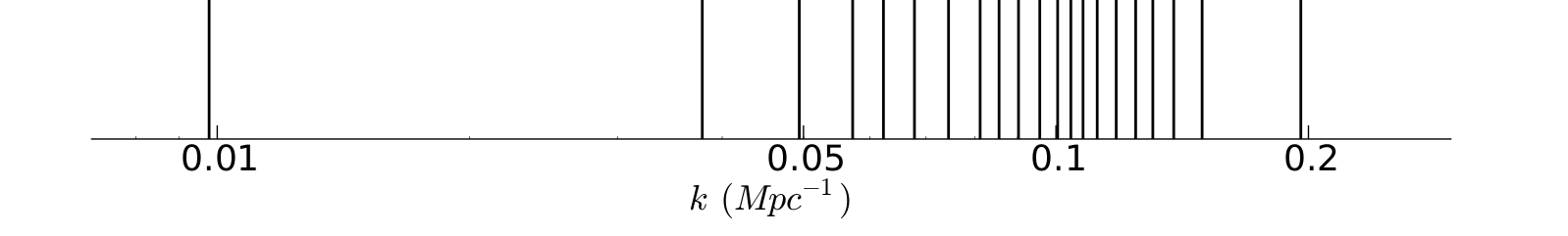}
\caption{
\baselineskip=0pt
For the partition into bins indicated at the bottom,
the images show the correlations between the bins using a greyscale.
100 bins were chosen so that the information in each bin as defined 
in the text is approximately equal. Each bin contains a 
$\chi ^2$ corresponding to a detection at about $100\sigma $ 
of power within that bin.
}
\label{BinCorr}
\end{center}
\end{figure}

We may choose bins of varying width so that each bin contains
roughly the same amount of information. The bottom of Fig.~\ref{BinCorr}
shows how the domain in wavenumber can be subdivided into
approximately 100 bins each containing about a detection at
$100\sigma $ (relative to the null hypothesis of a vanishing
power spectrum in that bin). 
The locations of the bin boundaries are shown in the lower panel
of the plot.
The upper panel of Fig.~\ref{BinCorr} 
shows using a greyscale density plot the correlations between
bins.

On smaller scales the story is more complicated. The precise
profile of the correlations near the diagonal is relevant 
for determining what information can be extracted concerning
small-scale structure in the power spectrum. 
In Fig.~\ref{transverse_xsec} the transverse profile
of the Fisher information density kernel is shown 
for some representative wavenumbers. 
The shape of the transverse profile is key to determining how much
information is available for reconstructing variations in
the primordial power spectrum that 
change rapidly with $k.$
A smooth broad profile acts to erase information on small scales in $k$
whereas a sharp narrow profile facilitates recovering information
concerning such rapid variations.  

Relative to a given experiment or collection of datasets, over what 
range in $k$ are
the parameters of the power spectrum, here $n_S$ and $\beta ,$
actually being measured? This is an important question
because the parametric model serves more as a local fitting function
for the power spectrum and does not derive from a fundamental
theory allowing us to shamelessly extrapolate the power law
as far as we want without trepidation. If we really knew that our 
power law ansatz were correct, this would not be a relevant
question. We would like to define $\kappa ^{low}$ and $\kappa ^{high}$
to characterize how far to the left and to the right
the bulk of the statistical information extends. Given 
the parameterization for the power spectrum 
\ba
P(k)=A(k/k_0)^{(n_S+\beta \ln (k/k_0))},
\ea
it follows that 
\ba
\ln P(k)=\ln A +
n_S \ln (k/k_0)+
\beta 
\ln ^2(k/k_0)
\ea
and to lowest order
\ba 
\frac{\delta P}{P_0}
=
\frac{\delta A}{A}
+(\delta n_S)(\kappa -\kappa _0)
+(\delta \beta ) (\kappa -\kappa _0)^2
\ea
where 
$\kappa =\ln k,$
$\kappa _0=\ln k_0,$
and $k_0$ is the so-called {\it pivot} or reference scale. 
Under a change of pivot scale $\kappa _0\to \hat \kappa _0,$ the parameters transform as 
\ba
A(\kappa _0)&\to & 
A(\hat \kappa _0)= 
\exp \left[ 
\ln A(\kappa _0)
+ n_S(\kappa _0)
 (\hat \kappa -\kappa _0)
+ \beta (\kappa _0)
(\hat \kappa _0-\kappa _0)^2
\right] ,\cr
n_S(\kappa _0)
&\to &
n_S(\hat \kappa _0)
=
n_S(\kappa _0)+
\beta (\kappa _0)
(\hat \kappa _0-\kappa _0),
\cr
\beta (\kappa _0) &\to &
\beta (\hat \kappa _0) =
\beta (\kappa _0)
\ea
where $\hat \kappa _0=\ln \hat k_0$ and 
$\kappa _0=\ln k_0$. 
It is sensible to choose $k_0$ so that the uncertainties in $A$ and
$n_S$ are uncorrelated. Otherwise, because of a lever arm effect,
the errors become large and highly correlated. 
Alternatively, the lack of correlation between $n_S$ and $\beta $ 
could have been chosen to
set $\kappa _0,$ but we have only one free parameter to adjust,
and both conditions place $k_0$ within what may be described
as the `center of mass' of the statistical information 
concerning the form of the primordial power spectrum for a 
specific experiment or a collection of datasets combining
multiple experiments. 

Having chosen a sensible pivot scale, we first consider a simplified situation with a diagonal Fisher
information density for the power spectrum so that
\ba
\chi ^2=
\int d\kappa ~
I_{diag}(\kappa ) 
[P_{model}(\kappa )- P_{data}(\kappa )]^2.
\ea
In this situation, the condition fixing the pivot scale is
\ba 
\int d\kappa ~ I_{diag}(\kappa ) (\kappa -\kappa _0)=0
\ea 
and the unmarginalized information (inverse of the variance
of the measurement with the other parameters fixed)
$n_S$ and $\beta $ are given by the two integrals
\ba 
I_{n_S}=
\int d\kappa ~ I_{diag}(\kappa ) (\kappa -\kappa _0)^2
\label{ns:info}
\ea
and 
\ba
I_{\beta }=
\int d\kappa ~ I_{diag}(\kappa ) (\kappa -\kappa _0)^4.
\label{beta:info}
\ea

For the Fisher density of an actual experiment, there are
always
correlations even if the information 
is strongly clustered near the diagonal and the integrals
in eqns.~(\ref{ns:info}) and (\ref{beta:info})
become double integrals. 
We modify
eqns.~(\ref{ns:info}) and (\ref{beta:info})
by limiting the integration over just one of the two
variables of integration. 
For the information concerning each variable, 
the cumulative information up to a certain $\hat \kappa $
is defined as follows, with the expression for $n_S$ given by
the integral
\ba
I_{n_S}(\hat \kappa )=
\int _{-\infty }^{\hat \kappa }d\kappa ~ 
\int _{-\infty }^{+\infty }d\kappa' ~ 
I(\kappa - \kappa'/2,\kappa +\kappa'/2) (\kappa -\kappa _0)^2,
\label{CumInfPSEqn}
\ea
and we define 
$\kappa _{n_S}^{low}$ and $\kappa _{n_S}^{high}$
as the locations of the the first and third quartiles, respectively, so that
$I_{n_S}(\kappa _{n_S}^{low})=(1/4)I_{n_S}$
and
$I_{n_S}(\kappa _{n_S}^{high})=(3/4)I_{n_S},$
and
$\kappa _{\beta}^{low}$ and $\kappa _{\beta}^{high}$
can be defined analogously.

\begin{figure}[htpb]
\begin{center}
\includegraphics[width=15cm]{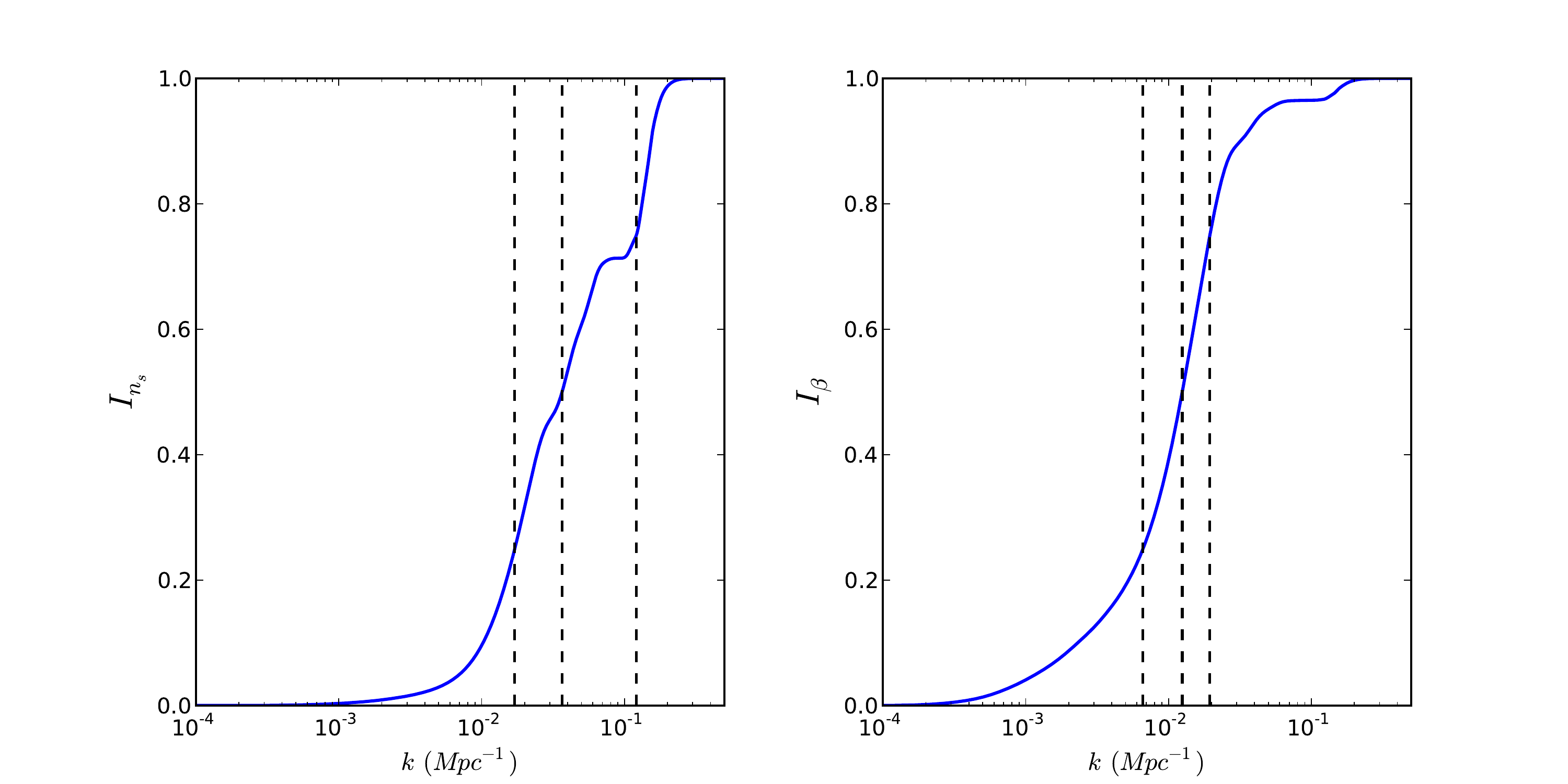}
\caption{\baselineskip=0pt
{\bf Cumulative information for $n_S$ and $\beta .$}
The left panel shows 
the cumulative information for $n_S$ as defined in eqn.~(\ref{CumInfPSEqn})
as well as the first, second, and third quartiles. The right panel shows
the analogous information for the running in the spectral index $\beta .$ }
\label{CumInfPS}
\end{center}
\end{figure}

The shape of the cumulative 
information for $n_S$ and $\beta $ are shown in Fig.~\ref{CumInfPS}
as well as the locations of the quartiles. (Note that when the matrix is not 
concentrated along the diagonal and especially in the presence of 
negative correlations assigning information to a specific $k$ no
longer makes sense and the above definitions become problematic.)

\clearpage
\section{Power Spectrum Reconstruction}

\subsection{Basic approach}

The previous section described the Fisher information density for the 
fractional variation 
\ba
f(k)=\delta P(k)/P_0(k)
\ea
of the primordial power 
spectrum relative to a reference model taken to be the 7-year WMAP best-fit model 
\cite{Komatsu:2010fb} already described in the previous section.
The Fisher 
information density $\hat I(k ,k^{\prime })$ was considered both with the vector 
$\betab $ containing the other cosmological parameters fixed---that is, assumed to
be determined extremely precisely from other non-CMB datasets---and with $\betab$ determined 
internally using the CMB data alone and hence uncertain. In the latter case we marginalize over the other 
cosmological parameters treating them as nuisance parameters with a flat (i.e., 
non-informative) prior probability distribution. 
In this section we consider how best to reconstruct the 
power spectrum---or more precisely $f(k)$---over the range of wavenumbers where the CMB provides 
sufficient information.

In principle, if the parameters (or perhaps bins) used to describe $f(k)$ are less numerous 
than the number of independent data points, we could reconstruct $f(k)$ without a 
regulator, or prior distribution concerning the smoothness of $f(k)$ in the form of a roughness 
penalty. However since we want to avoid a parametric model as well as artifacts from 
bin boundaries, we instead introduce a roughness penalty. The Fisher information matrix for the primordial power 
spectrum takes the general form
\ba
I_{ff}= (X^T)_{fc} I_{cc} X_{cf}+\lambda R_{ff}.
\ea
In the quadratic approximation, minus twice the log likelihood is given by $\chi ^2=\f^{T} 
I_{ff} \f$. Here $I_{cc}$ is Fisher matrix for the angular CMB power spectrum. In order to 
provide as compact a notation as possible, we write our expressions as if we were dealing 
with finite-dimensional matrices and vectors, even though the index $f$ represents continuous variable and 
$c$ is a discrete index. Here $\lambda R_{ff}$ represents a regulator whose eigenvalues 
should very nearly vanish for slow variations in the power spectrum but be 
large (relative to the eigenvalues of the Fisher matrix) for modes where $f(k)$ wiggles 
rapidly. A commonly used but by no means unique choice for this regulator is the 
quadratic form
\ba
\lambda R(f)=
\lambda 
{\bf f}^T
R_{ff}
{\bf f}
=\lambda \int d\kappa
\left(
\frac{
\partial ^2f(\kappa)
}{
\partial \kappa^2
}
\right) ^2,
\label{XXA}
\ea
which is a measure of how much the function $f(k)$ deviates from a straight line. Under 
this regulator, to linear order a change in the amplitude $A_{s}$ or spectral tilt $n_S$ 
suffers no penalty at all, but rapid variations in the running of the spectral index, or 
a large curvature, are penalized by an amount proportional to the metaparameter $\lambda$. 
The object is to penalize functions that are too rough through a judicious choice for 
$\lambda$.

Above we have assumed that the other {\it cosmological parameters}, or
elements of the vector $\betab ,$ have been fixed. If instead these are determined 
internally using the CMB data, the expression for the Fisher information for the primordial 
power spectrum after marginalization is modified to 
\ba 
I_{ff}^{(marg,flat)}=
(X^T)_{fc} 
\left[
I_{cc}-
(I_{cc}X_{c\beta })
(X^T_{\beta c}I_{cc}X_{c\beta })^{-1}
(X^T_{\beta c}I_{cc})
\right]
X_{cf},
\label{MargFisher}
\ea
and in the presence of additional information on $\betab$ (for example, by incorporating 
other external non-CMB data sets) expressed by the Fisher matrix $I_{\beta \beta },$ this 
expression modifies to
\ba 
I_{ff}^{(marg,prior)}=
(X^T)_{fc} 
\left[
I_{cc}-
(I_{cc}X_{c\beta })
(X^T_{\beta c}I_{cc}X_{c\beta }+I_{\beta \beta })^{-1}
(X^T_{\beta c}I_{cc})
\right]
X_{cf} .
\label{MargFisher_w_prior}
\ea
The equations here assume a Gaussian likelihood as well as Gaussian prior information. The 
most significant non-Gaussianity arises from the angular power spectrum likelihood at very 
low $\ell $ and from nonlinearity in the change of the angular power spectrum as a function 
of variations in the cosmological parameters (i.e., the components of the vector $\betab$). 
Supposing Gaussianity allows simple analytic expressions to be obtained. In practice the 
distributions are slightly non-Gaussian and must be explored using MCMC or some other 
numerical sampling method. The Gaussian analysis, however, is invaluable for gaining 
intuition about how to select a good regulator, but the solutions obtained should 
subsequently be validated by simulations taking into account the 
non-Gaussianity of the regulated likelihood.

\begin{figure}[htb]
\begin{center}
\includegraphics[scale=0.5]{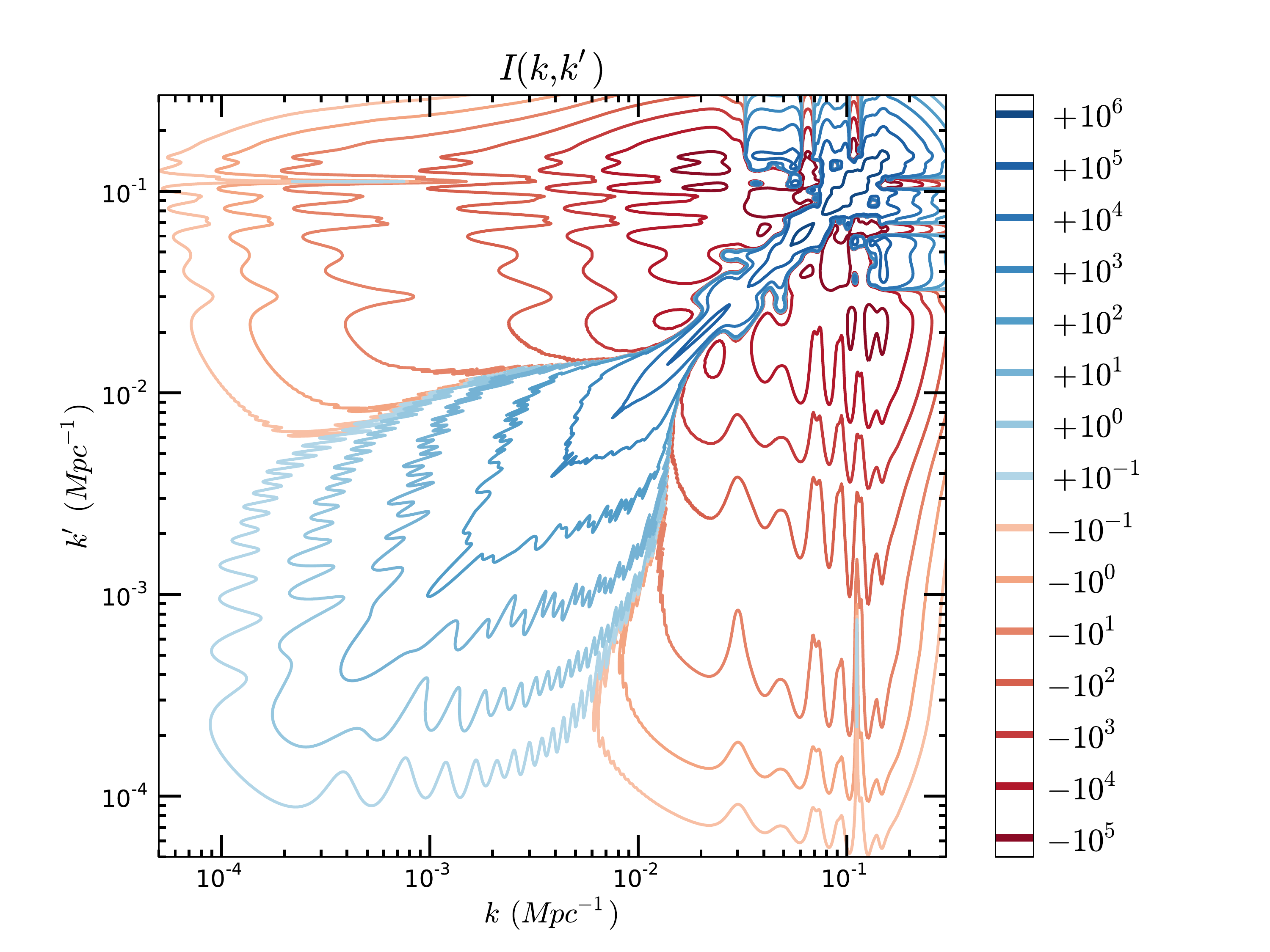}
\caption{{\bf Fisher density contour plot}
(after marginalization over the cosmological parameters).}
\label{Iij_marg_smoothed}
\end{center}
\end{figure}

\begin{figure}[htb]
\begin{center}
\includegraphics[width=15cm]{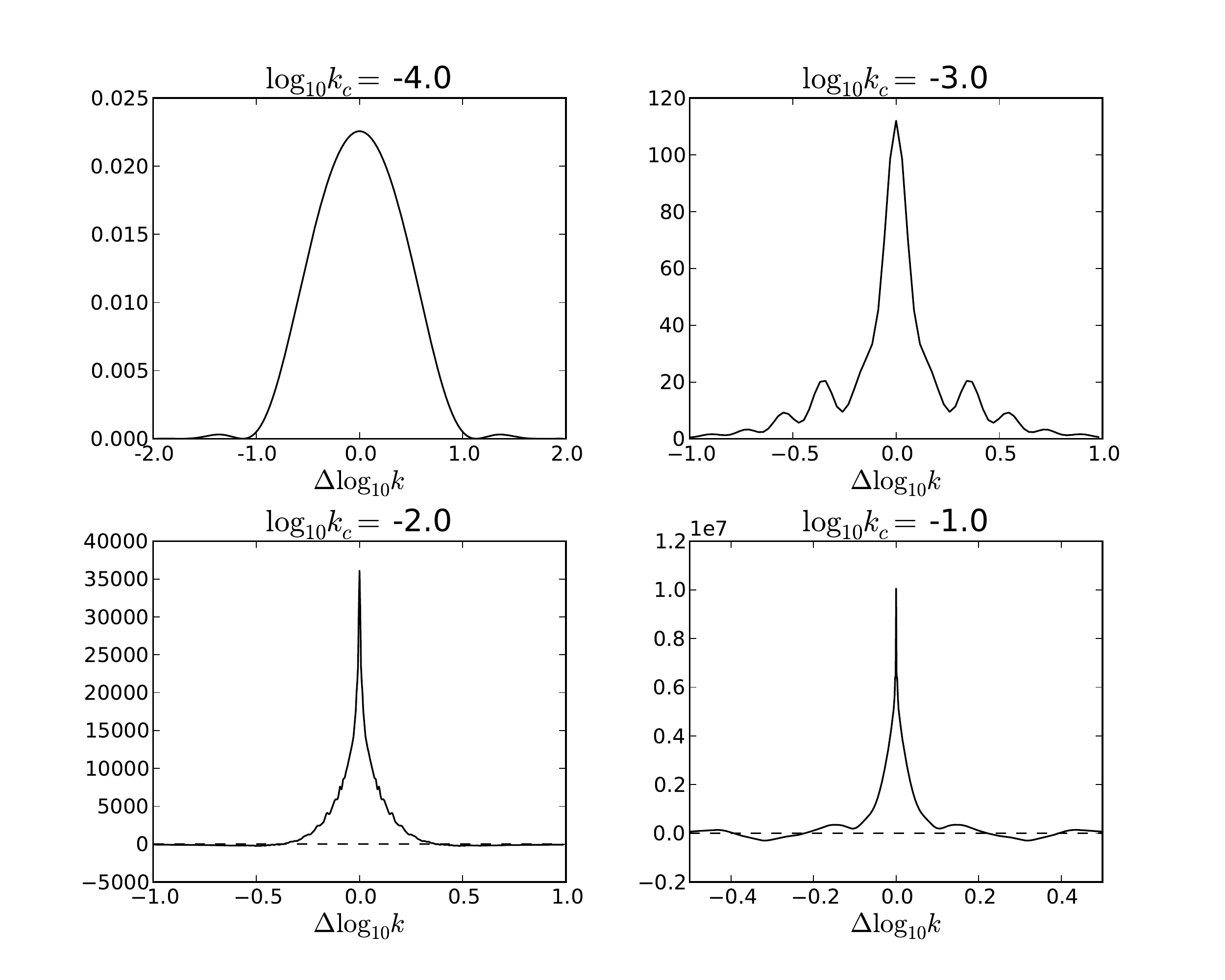}
\caption{
\baselineskip=0pt 
{\bf Cross sections of the Fisher information density along 
transverse direction
(with cosmological parameters uncertain and determined from CMB data alone)} centered on the diagonal.
The $k_{c}$ indicate the diagonal crossings.
Compare these plots to Fig.~\ref{transverse_xsec}. 
At large scales ($\log_{10} k_{c} = -4$ and $\log_{10} k_{c} = -3$) 
the kernel is virtually identical in both figures, while at small scales 
the behavior changes depending on whether the cosmological parameters 
are fixed or internally determined with uncertainty.}
\label{transverse_xsec_marg}
\end{center}
\end{figure}

\begin{figure}[htb]
\begin{center}
\includegraphics[scale=0.45]{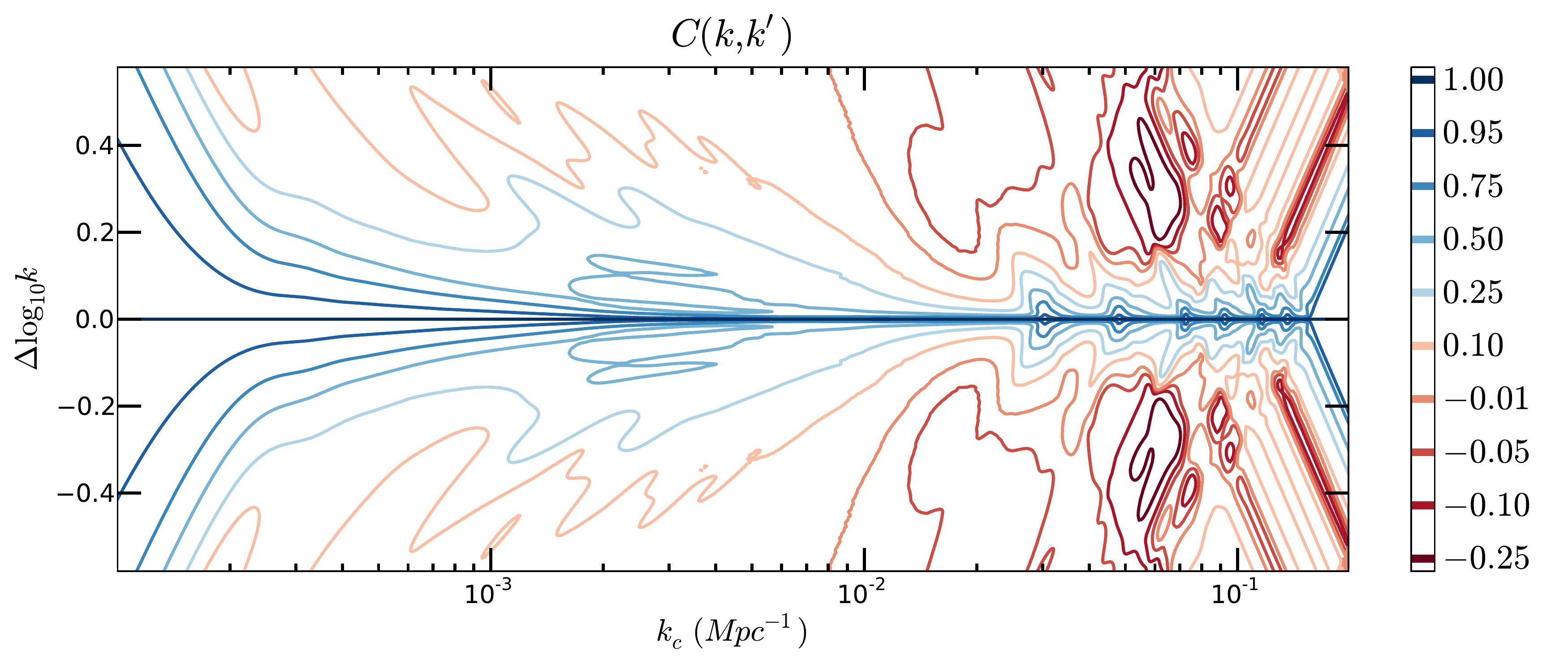}
\caption{
\baselineskip=0pt 
{\bf Correlation after marginalization over the uncertain
cosmological parameters determined from CMB alone.} Note that there is now much more correlation
between very low and very high $k$.}
\label{corr_marg_smoothed}
\end{center}
\end{figure}

\begin{figure}[!h]
\begin{center}
\includegraphics[scale=.4]{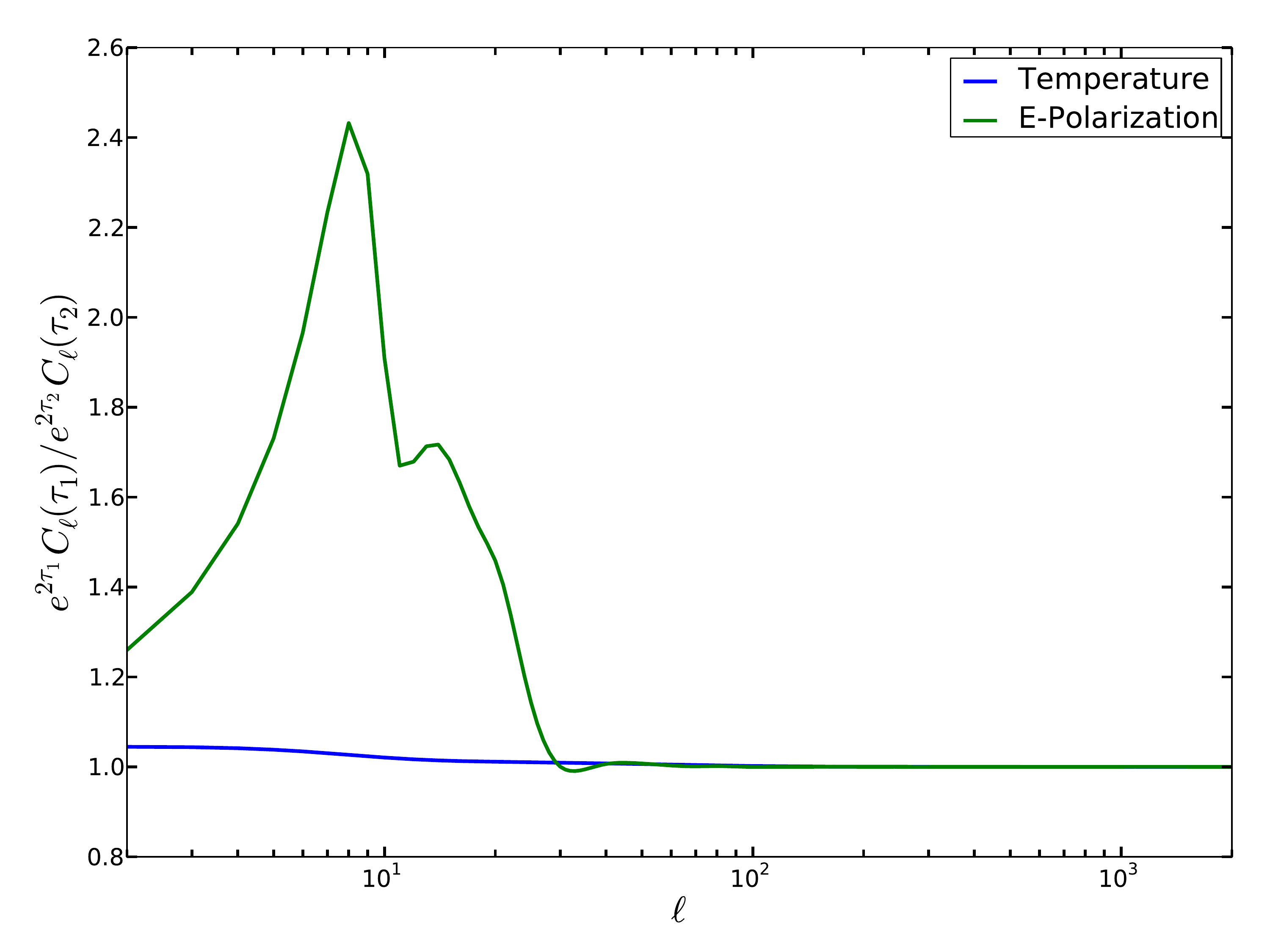}
\caption{
\baselineskip=0pt
Ratio of the temperature and E-polarization $C_{\ell}$'s at two different values of the 
optical depth: $\tau_{1}=0.1$, $\tau_{2} = 0.08$. 
For $\ell \gtorder 30,$ the shapes differ by less than 1\%, and for $\ell \gtorder 100,$ 
less than $0.1\%.$
}
\label{tau_cl_comparison}
\end{center}
\end{figure}

Fig.~\ref{Iij_marg_smoothed} shows the Fisher density 
when $h,$ $\omega _b,$ and $\omega _c$ are uncertain and constrained only
by the CMB data, and Fig.~\ref{transverse_xsec_marg} shows some transverse profiles.
These should be compared with the analogous plots in Figs.~\ref{contour_Kern} and 
\ref{transverse_xsec} of the unmarginalized Fisher density, which are almost identical at 
small $k$, but have significantly different behavior at high $k$. Marginalization over the 
cosmological parameters increases the inter-$k$ correlations, as seen
in Fig.~\ref{corr_marg_smoothed}. In marginalizing over the cosmological parameters, we 
have not included the reionization 
optical depth $\tau$, which is doubtless a relevant parameter. This is because except at 
very low $\ell$, the reionization optical depth
$\tau$ and the overall amplitude of the power spectrum $A_{s}$ are 
completely degenerate. It is the combination $A_{s} e^{-2 \tau}$ that normalizes the overall 
amplitude but not the shape of the CMB sky power spectrum. Fig.~\ref{tau_cl_comparison} 
shows the deviations to this statement at low $\ell$. The deviation is 
greatest at low $\ell$, especially for the E-mode polarization. 
The data from the very first multipoles can be used to determine $\tau$, and then to determine the 
overall normalization.

We now return to the problem of choosing the metaparameter $\lambda $ of the regulator. On the 
one hand, if $\lambda $ is too small, the reconstruction is dominated by noise, seeded by 
the cosmic variance of a specific sky realization and by random instrument noise on smaller 
scales. In this situation, the bias of the reconstruction will be small but its variance will
be large. On the other hand, 
if $\lambda $ is too large, real features in the primordial power spectrum, especially 
those that are sharp or highly localized, will be smoothed over and distorted. This is {\it 
bias} accompanied by very little {\it variance.} Finding the right $\lambda $ amounts to 
making the right compromise between bias and variance, both of which should be minimized,
resulting in conflicting requirements that cannot be satisfied simultaneously.

The effect of the regulator may be described using the language of Bayesian statistics, as 
a prior distribution over a function space. While this description provides useful 
insight, the resulting smoothness prior does not correspond to any actual prior 
belief concerning the theoretical power spectra. Rather the prior is chosen after the fact and 
is informed primarily by the characteristics of the experiment and the information provided 
by the data.

Automatic procedures can be derived for choosing the right parameters for the regulator. In 
the course of this study we tried using the Aikake Information Criterion (AIC) \cite{Aikake1974} to 
choose $\lambda $ as well as criteria penalizing model complexity more strongly inspired by 
the BIC (Bayesian Information Criterion) \cite{Schwarz1978}. However, while these criteria may offer some 
theoretical rational for selecting $\lambda ,$ they do not provide any insight into what 
kind of features in the power spectrum can and cannot be recovered.

The problem of choosing a regulator to recover the primordial power 
spectrum necessarily involves some degree of arbitrariness. The approach adopted here  is 
pragmatic. We construct a family of model features to be 
reconstructed and then test various regulators with a range of smoothing parameter choices.
The final product puts forth a reconstruction procedure whose properties have been well 
characterized by showing what features can and cannot be recovered given the experiment and 
at what statistical significance.

The functional form of the regulator adopted in eqn.~(\ref{XXA}) is somewhat arbitrary. 
This form could be justified if there were a translation invariance in the variable $\kappa 
=\ln k.$ The discussion of the previous section showed that the CMB provides 
stringent constraints on $f(k)$ over a window spanning approximately two decades in $k$, 
but outside this window $f(k)$ is hardly constrained. In fact, for $k$ well below this 
window---that is, for modes whose wavelengths are superhorizon today---we measure only the local 
curvature in the geometry and have no way of deducing the precise wavelengths of the 
long-wavelength modes giving rise to this curvature, which may be approximated 
within the present horizon using a 
quadratic approximation. This situation implies that only an integral over the very low-$\ell $ power 
spectrum is constrained by the CMB data. This fact is evident from the correlation structure of 
the Fisher matrix at low-$k$ (Figs.~\ref{Corr:Fig}, \ref{corr_marg_smoothed}), which shows that
the dimensionless correlation is almost one in the corner.  
Because of 
the sharpness of the centrifugal barrier in the differential equation for the spherical 
Bessel functions, the very low-$k$ power spectrum affects only the quadrupole and has almost 
no impact on the higher-$\ell $ multipoles. The impact of the high-$\ell $ part of the 
power spectrum lying above this window is more complicated because at large arguments the 
spherical Bessel functions oscillate rapidly and fall off slowly according to a power law 
that does not depend on $\ell$.

This situation suggests that more regulation is needed outside this window than inside, 
suggesting the following generalization of eqn.~(\ref{XXA})
\ba
\lambda 
{\bf f}^T
R_{ff}
{\bf f}
=\int d\kappa
\rho (\kappa)
\left(
\frac{
\partial ^2f(\kappa)
}{
\partial \kappa^2
}
\right) ^2 ,
\label{XXB}
\ea
yet this proposal introduces additional arbitrariness in the choice of $\rho$. In order to 
deal with the region of the power spectrum that is hardly constrained by the CMB data, we 
found it necessary to modify the regulator by adding a term
pushing the reconstructed spectrum toward the fiducial model. Thus the 
regulator becomes
\ba
{\bf f}^T
R_{ff}(\lambda,\alpha)
{\bf f}
=
\lambda \int 
d\kappa~
\left(
\frac{
\partial^2 f(\kappa)
}{
\partial \kappa^2
}
\right) ^2
+\alpha \int _{-\infty }^{\kappa _{min}}
d\kappa~f^2(\kappa)
+\alpha \int ^{+\infty }_{\kappa _{max}}
d\kappa~f^2(\kappa)
\label{Priors}
\ea
and now has four adjustable parameters. The last two terms 
constitute an endpoint fixing regulator. Without these terms,
the smoothing regulator favors a linear PPS on the far left and
the far right where the PPS is almost
unconstrained by the CMB data. This behavior leads 
to spurious deviations from the
fiducial model in the reconstructions.

The reconstruction process may be characterized as follows. Let $f_{in}$ be the 
input fractional variation of the PPS containing a number of features
to be reconstructed.  The recovered fractional variation is  
\ba 
f_{recov}= \bar f_{recov}+\delta f, \ea 
which is decomposed into a mean component $\bar f_{recov}$ and a random 
fluctuating component (of vanishing mean) $\delta f,$ whose
covariance is given by
\ba 
\left< \delta f ~\delta f \right> =
\Bigl( I+R(\lambda ,\alpha ) \Bigr) ^{-1} 
I
\Bigl (I+R(\lambda ,\alpha ) \Bigr) ^{-1} 
\label{covariancef} 
\ea 
and its mean is 
\ba 
\bar f_{recov} 
&=& [I_{ff}+R_{ff}(\lambda , \alpha )]^{-1}I_{ff} f_{in} = Af_{in} 
\label{reconEq} 
\ea 
The reconstruction operator $A$ is a smoothing operator, which diminishes noise in 
$f_{in}$ from cosmic variance and detector fluctuations by averaging or smoothing, but also introduces a bias into the
reconstruction $\bar f_{recov}$. 
This bias is unavoidable since some degree of 
smoothing is needed to avoid overfitting the data. 

For a numerical implementation, we must discretize $f(k)$ by restricting to a 
finite-dimensional function space. We use a function
space sufficiently large so that the results are insensitive to
the dimension used. In the interest of numerical stability, it is advantageous to reduce 
the dimensionality of the function space as much as possible.
We found that cubic $B$-splines are well suited for avoiding 
discretization artifacts 
while keeping the number of control points small. We expand as 
\ba
f(\kappa) = \sum_{i} f_{i} B^{i}(\kappa) .
\ea
where $B^{i}(\kappa)$ are cubic $B$-spline basis functions.
We work in the continuum limit where the roughness penalty suppresses variations on 
scales close to the control point spacing.  Throughout a grid equally spaced 
in $\kappa = \ln k$ with $\Delta \kappa = 0.01$ is used.

\subsection{Reconstruction bias with other cosmological parameters fixed}

\begin{figure}
\begin{center} 
\includegraphics[scale=.9]{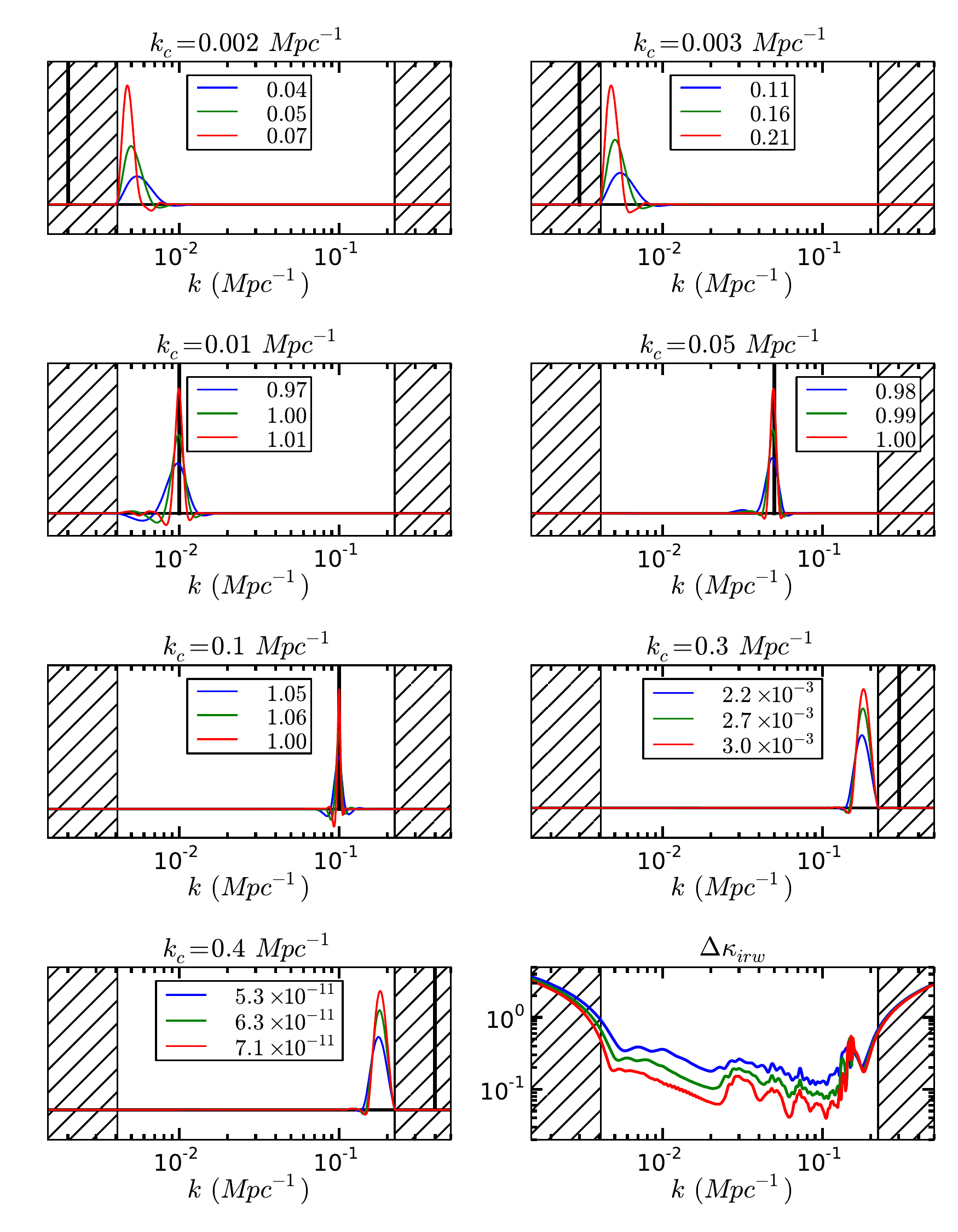} 
\caption{
\baselineskip=0pt
{\bf Impulse response of reconstruction operator $A$} (case with 
other cosmological parameters fixed).
The curves correspond to 
$\lambda = 10^{4}$ (blue), $\lambda = 10^{3}$ (green) and 
$\lambda = 10^{2}$ (red). 
In the shaded regions on the far left and right 
(where $\kappa <\kappa_{low} = -5.5$ or $\kappa >\kappa_{high} = -1.5$)
an endpoint fixing prior (with $\alpha = 10^{4}$) has been applied,
 pushing the reconstruction $f$ to zero there. The numbers in the boxes
indicate the area under the curves of the reconstruction kernel, which 
ideally should be as close as possible to one, the area under the delta
function input. The values in the boxes in the two last plots show that
there is little leakage from variations in the shaded regions.
The lower right plot shows the impulse response 
width as a function of feature position.
}
\label{ImpluseResponse}
\end{center}
\end{figure}

In this subsection we examine the mean reconstruction error when the other cosmological parameters 
(i.e. $H_0,$ $\omega _b,$ $\omega _c$, $\tau$, etc.)
are fixed.
In the next subsection we generalize taking into account uncertainties in these parameters and the 
following section examines the random error and questions of statistical significance.
Fig.~\ref{ImpluseResponse} plots the impulse 
(i.e., $\delta $-function input)
response of $A$ 
without marginalizing over the cosmological parameters. The action of the reconstruction
operator $A$ on an impulse 
spreads a sharp localized input feature. Mathematically this is due to the smoothing 
regulator, which smears the variations in the power spectrum 
where its second derivative is large. 

The reconstruction operator introduces a bias in the reconstructed signal that underestimates
the height of features. Since the process is linear, a full description of the bias takes the
form of the two-dimensional kernel, of which some representative cross sections are shown
in Fig.~\ref{ImpluseResponse}. However two quantities provide a succinct partial
characterization of the kernel: the {\it impulse response width,} 
defined as the standard deviation or square root of the second moment about the mean 
as a function of $k_{in};$ and the area under the curve, whose deviation from one 
indicates how much the amplitude of broad features is distorted.

The limits $\kappa _{low}$ and $\kappa _{high}$ of
the endpoint fixing regulator in eqn.~(\ref{Priors}),
beyond which the spectrum is effectively constrained to coincide with the fiducial model,
single out an interval of $k$ space where to search for features. 
A concern is that deviations from the fiducial spectrum
beyond these limits (where we cannot
measure the power spectrum) could result in spurious features inside the free interval.
We found that this not to be 
a problem when the other cosmological parameters were fixed. 
Several of the plots in Fig.~\ref{ImpluseResponse} 
show the impulse response where the input is located in the 
fixed regions and we observe that except for some small
artifacts near the boundary
of a small integrated area, 
there are no artifacts in the interior of the free region.

\subsection{Reconstruction bias with uncertainty in cosmological parameters}

The situation studied in the previous subsection where the other
cosmological parameters not associated with the PPS
are assumed known with absolute precision is
idealized. In practice, whether these parameters are
determined internally from the CMB data alone or whether additional
non-CMB data is also used (acting as a prior on these parameters),
some uncertainty in these additional parameters is always present. This
uncertainty degrades the quality of the power spectrum
reconstruction, because variations in the power spectrum
can be traded against changes in these parameters. This subsection
describes the properties of the mean reconstruction in the
presence of these uncertainties, with and without external information. 
The next section treats the noise properties of the reconstruction.

We found that for the case where the other cosmological parameters were
determined from the CMB data alone,
in order to prevent possible features in the fixed region
from introducing spurious features in the free region and spurious
shifts in the cosmological parameters, it is necessary to choose a
wide interval $[\kappa _{low},\kappa _{high}]= [10^{-4} , 0.3] \textrm{ 
Mpc}^{-1} $  for the free region.  $k_{low}$ here 
corresponds roughly to the current horizon scale ($k_{low} \sim 1/\eta_{0}$ where 
$\eta_{0}$ is the distance to the last scattering surface).
$k_{high}$ is slightly above the $k$ corresponding to $\ell _{max}.$ 
The situation contrasts with the case of the previous subsection
where the cosmological parameters
are fixed, for which $[\kappa _{low},\kappa _{high}]$ can be chosen freely
without introducing artifacts.  
For the remainder of the paper we set $\alpha=10^{4}$ and define the free
region as $[10^{-4} , 0.3] \textrm{ Mpc}^{-1}.$ 

\begin{figure}[htb]
\begin{center}
\includegraphics[scale=.6]{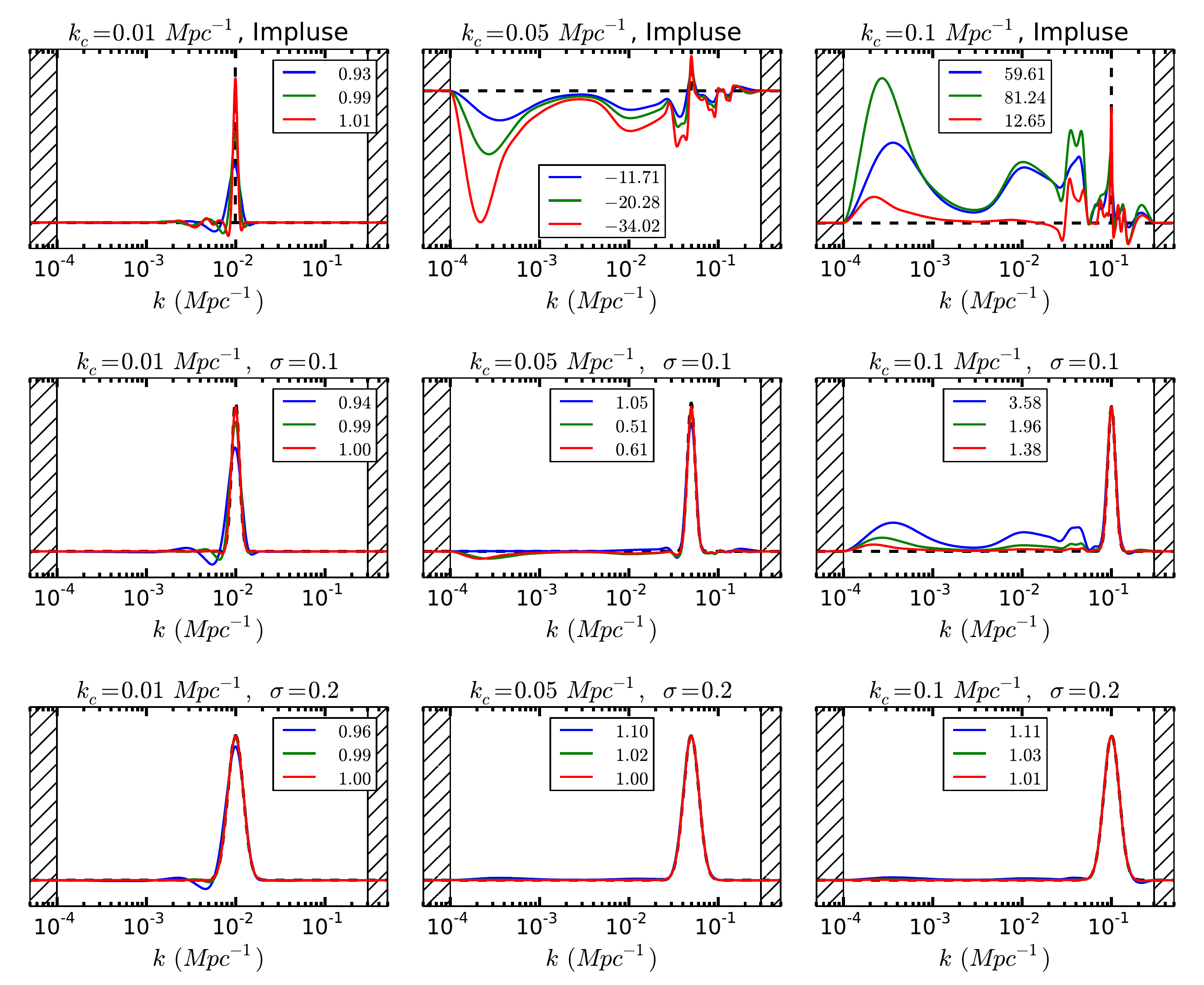}
\caption{
\baselineskip=0pt
{\bf Smoothing Response to impulse and Gaussian inputs}
(with marginalization over the other cosmological parameters).
The values of the prior 
parameters and their color mapping are the same as in Fig.~\ref{ImpluseResponse}. The
endpoint fixing regulator limits here are $k_{low} = 10^{-4}$ Mpc${}^{-1}$ and $k_{high} = 0.3$ 
Mpc${}^{-1}$.}
\label{marg_response_array_w_area}
\end{center}
\end{figure}

\begin{figure}[htb] \begin{center} 
\includegraphics[scale=.4]{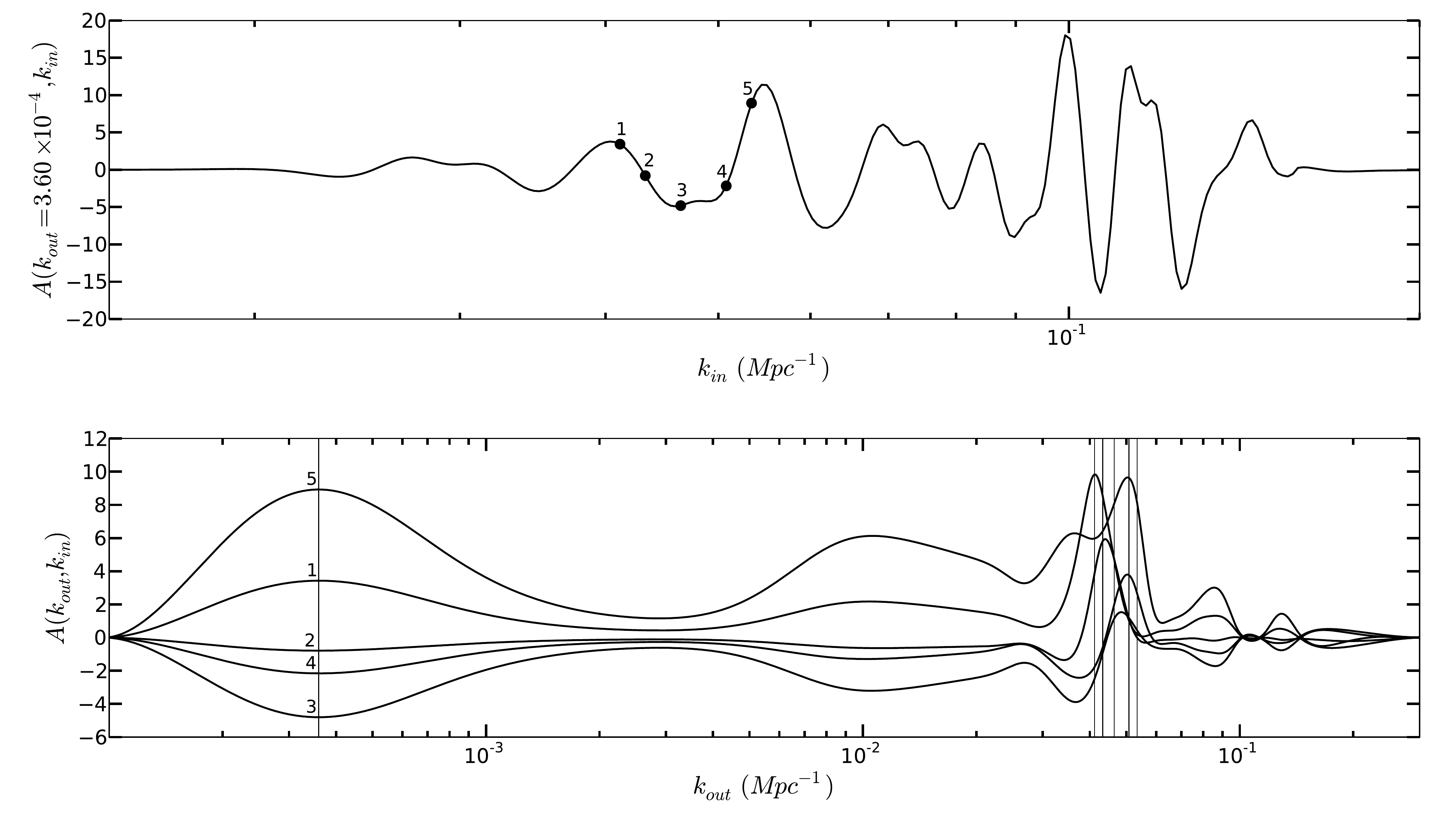} 
\caption{ \baselineskip=0pt 
{\bf Rapid oscillations in the marginalized smoothing operator 
$A(k_{out},k_{in})$.} The bottom plot shows $A(k_{out},k_{in})$ versus $k_{out}$ for 
different $k_{in}$. The black vertical bars on the right 
indicate five values of $k_{in}$ for which $A(k_{out},k_{in})$ is plotted. We observe that 
the shape of the tail of the kernel remains approximately fixed but its overall amplitude 
undergoes oscillations as $k_{in}$ is varied. The top figure shows 
$A(k_{out},k_{in})$ as a function of $k_{in}$ with $k_{out}$ fixed at the value of $k$ 
corresponding to the position of the single vertical line on the left in the bottom plot. 
The points show the $k_{in}$ values for the curves of the bottom plot.} 
\label{Aopertor_plot} 
\end{center} 
\end{figure}

The top row of Fig.~\ref{marg_response_array_w_area} shows that the 
reconstruction operator badly distorts an 
infinitely narrow impulse input in the case where 
$h,$ $\omega _b,$ and $\omega _c$ are uncertain and constrained only by the CMB data.
This 
distortion is due to trading variations in the cosmological parameters
against variations in the power spectrum, leading to a reconstruction
kernel having long oscillating tails in the absence of additional 
non-CMB information, as in the case plotted here.
Fig.~\ref{marg_response_array_w_area} also shows the response to a Gaussian input
profile of varying width as plotted in the bottom two rows. 
The bad behavior in the tails of the reconstruction 
kernel almost disappears when the kernel is applied to a
broadened input.
Fig.~\ref{Aopertor_plot} shows that in 
the region of $k$-space where the bias is most extreme, the response of an impulse 
away from the central peak oscillates rapidly as the location $k_{c}$ of the input
impulse is varied. Although the 
amplitude of the response in the tail oscillates, its shape remains roughly constant
as shown in the figure.
This is why the oscillations in the tail
interfere destructively almost canceling each other
when the impulse input is replaced with a broadened Gaussian input.

Allowing for uncertainty in $h$, $\omega_{c}$ and $\omega _b$, not only modifies the 
reconstruction kernel, but also causes features in the power spectrum 
to displace the maximum likelihood cosmological parameters. In the absence of 
a regulator the mean of this effect would vanish, but the mean is nonzero for $\lambda >0.$
By expanding the transfer function to first order around the input cosmological 
parameters, we can express the change in the $C_{\ell}$'s as a linear function of the 
change in the PPS and the change in the cosmological parameters $\delta \betab$:
\begin{gather}
\delta C_{\ell}   = T_{c f}f + T_{c \beta} \delta \betab .
\label{LinApproxCl}
\end{gather}
If we treat $(f,\delta \betab)$ as a single parameter vector, we can write an extended 
reconstruction operator $A$ that acts on the space of $(f,\delta \betab)$, which we can write in 
block form as
\begin{gather}
A = \left[
\begin{array}{cc}
A_{ff} & A_{f\beta}\\
A_{\beta f} & A_{\beta \beta}
\end{array}
\right]
\end{gather}
$A$ is defined by inserting $\hat I$ from 
eqns.~(\ref{MargFisher}) or (\ref{MargFisher_w_prior}) into eqn.~(\ref{reconEq}).
Due to the off-diagonal block $A_{\beta f}$, a feature in the PPS will in general shift the values of the 
cosmological parameters.

\begin{figure}[!h]
\begin{center}
\includegraphics[scale=.4]{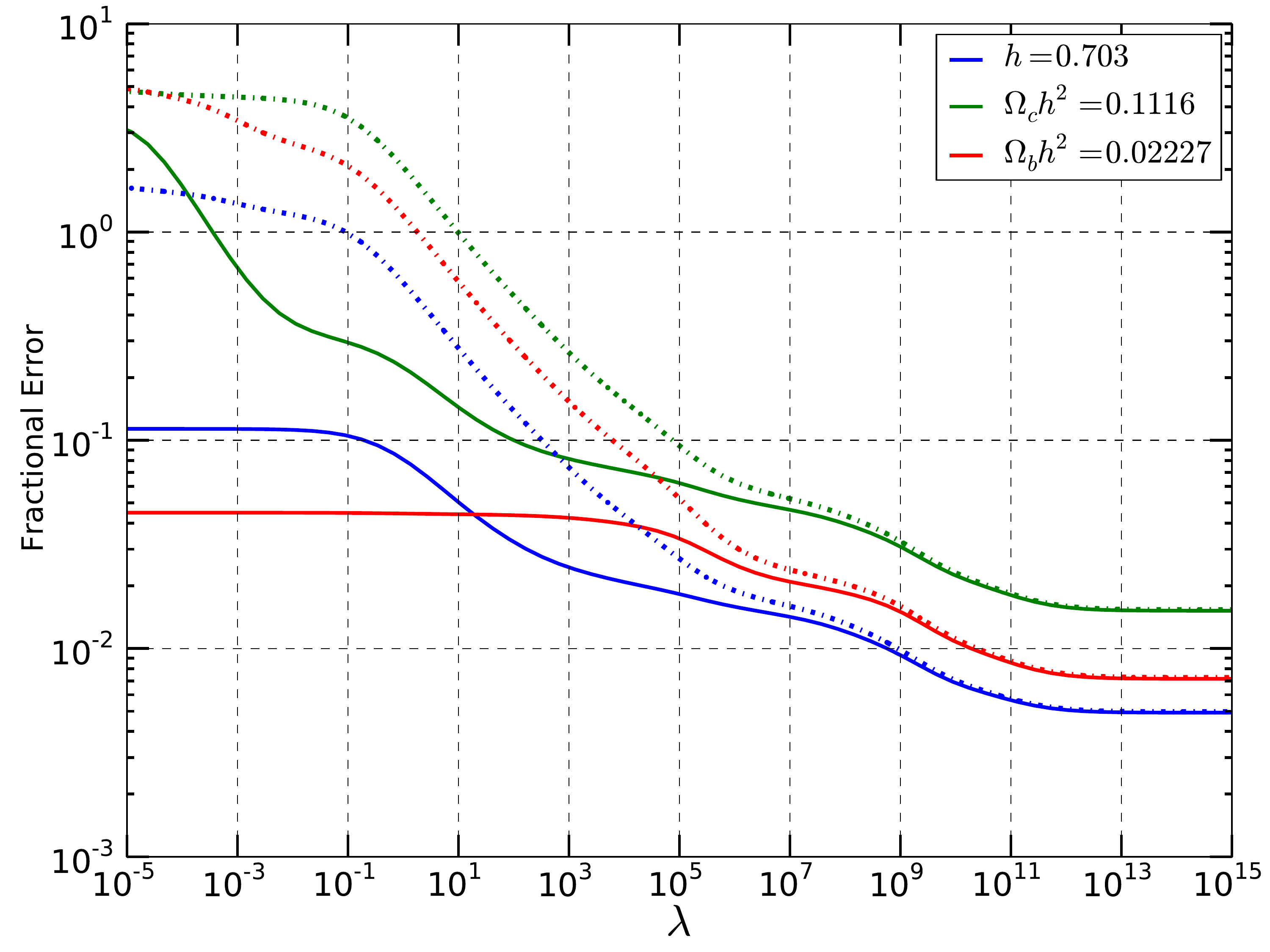}
\caption{
\baselineskip=0pt
{\bf Fractional error in cosmological parameters as a 
function of the smoothing parameter $\lambda$.} The solid curves show the errors with 
Gaussian priors on $h$ and $\Omega_{b} h^{2}$, while the dashed curves show the errors 
without priors. At extremely large values of the smoothing parameter ($\lambda>10^{12}$), 
the form of the feature is restricted to a straight line, reducing the effective degrees 
of freedom in $f(k)$ to two (a magnitude and tilt). In this case the error on the 
cosmological parameters is at its absolute minimum.
At the other extreme, with $\lambda\ltorder 10^{-1}$ the form of the PPS is almost totally 
unconstrained and the constraints on the cosmological parameters are those from the
external data. 
}
\label{cosmo_err}
\end{center}
\end{figure}

For small $\lambda $ it is useful to include other data to prevent the other parameters
from taking unreasonable values.
The constraints from external data used here are
$h = 0.72 \pm 0.08$ from \cite{0004-637X-553-1-47} (HST key project)
and $\Omega_{b} h^{2} 
= (2.2 \pm 0.1) \times 10^{-2}$ from \cite{omeara2001,2008MNRAS.391.1499P}
(measured deuterium abundances), but
$\Omega_{c} h^{2}$ is left unconstrained  by the external data. These constraints are implemented as
Gaussian priors in our calculations.
Fig.~\ref{cosmo_err} shows that in the absence of external data, 
for small values of $\lambda $ the determination of the cosmological parameters degrades
considerably, contradicting what is known from external data.
With external data constraining $h$ and $\omega _b$
the situation is remedied substantially, although  
for interesting values of $\lambda ,$ up to 20\% errors in $\Omega_{c} h^{2}$ persist.

\begin{figure}[htb]
\begin{center}
\includegraphics[scale=.6]{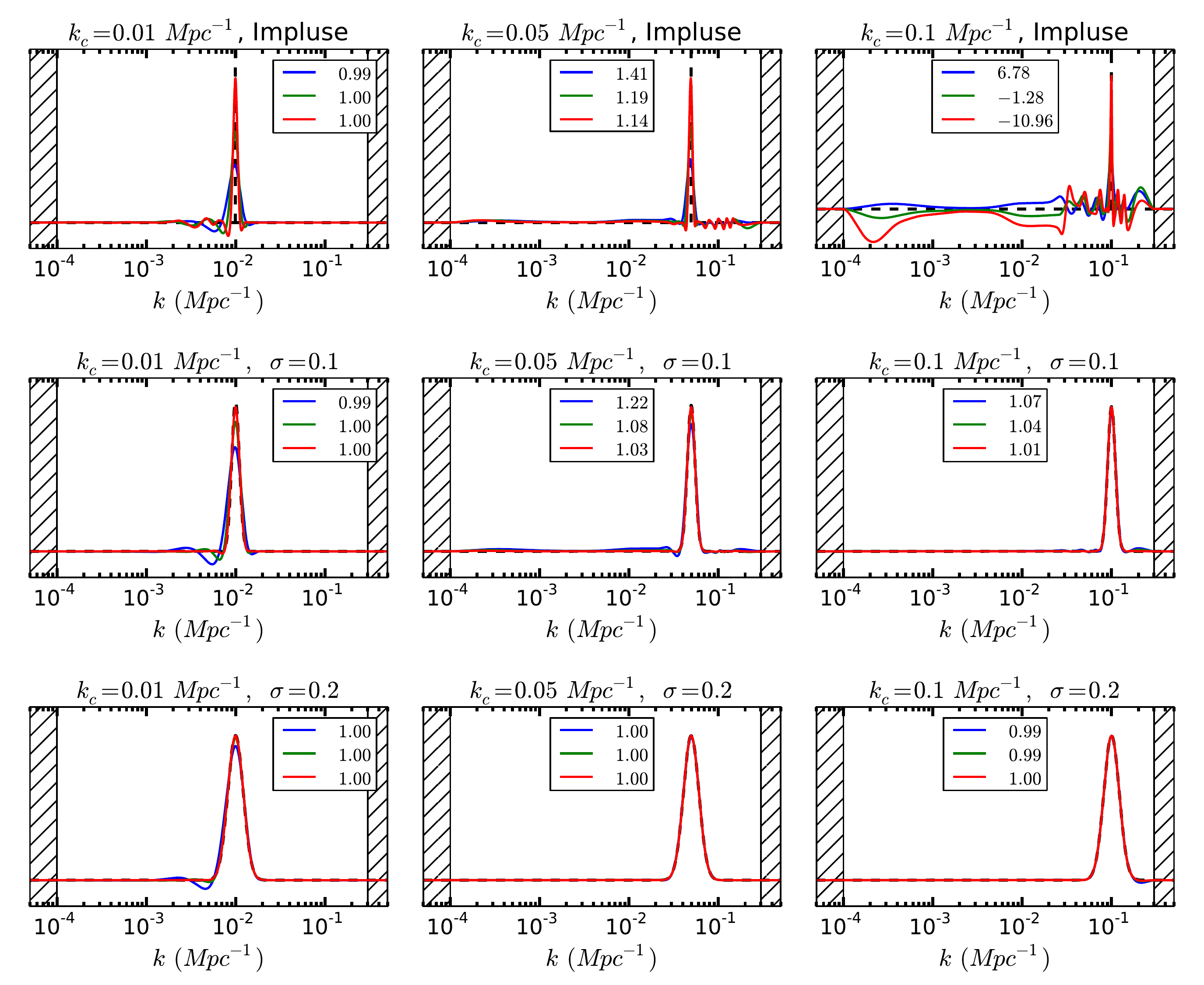}
\caption{
\baselineskip=0pt
{\bf Smoothing response to impulse and Gaussian inputs}
(with marginalization and external priors over the other cosmological parameters).
The values of the prior 
parameters and their color mapping are the same as in Fig.~\ref{ImpluseResponse}. The
endpoint fixing regulator limits here are $k_{low} = 10^{-4}$ Mpc${}^{-1}$ and $k_{high} = 0.3$ 
Mpc${}^{-1}$.}
\label{marg_response_array_extern_prior_w_area}
\end{center}
\end{figure}

With external datasets included, the deformation of a signal due to the 
reconstruction operator is lessened. In Fig.~\ref{marg_response_array_extern_prior_w_area}, 
the impulse and Gaussian response of the reconstruction operator is shown. 
Compared to Fig.~\ref{marg_response_array_w_area}, the response is better behaved 
although still not as clean as the response when the cosmological parameters have been fixed 
(Fig.~\ref{ImpluseResponse}). This improvement is due to the decreased uncertainty in the 
cosmological parameters, which in turn lessens the need for the PPS to compensate for a 
shift in the cosmological parameters.

Figs.~\ref{marg_response_array_w_area} and \ref{marg_response_array_extern_prior_w_area}
demonstrates that features that are too narrow cannot be reconstructed faithfully
because of confusion with variations of the other cosmological parameters. We
define the {\it minimum reconstructible width} $\Delta \kappa _{mrw}$ to describe this situation
quantitatively. The quality of a reconstruction can be characterized by 
the mean square error, and setting the somewhat arbitrary threshold here to $b=10^{-2},$ we 
deem a reconstruction to be satisfactory when
\begin{gather}
\int d \kappa \left| 
A f_{\kappa_{c},\sigma }(\kappa) - f_{\kappa_{c},\sigma }(\kappa)\right|^{2} 
\le
b \int d\kappa \left|f_{\kappa_{c},\sigma }(\kappa)\right|^{2}.
\label{biasCriterion}
\end{gather}
Here $f_{\kappa_{c},\sigma }$ is a Gaussian centered at $\kappa_{c}$ with 
standard deviation $\sigma .$ For a given $\kappa _c,$ 
the minimum reconstructible width $\Delta \kappa _{mrw}$ is
defined as the smallest $\sigma $ such that the above condition still holds.

\begin{figure}[htb]
\begin{center}
\includegraphics[scale=.5]{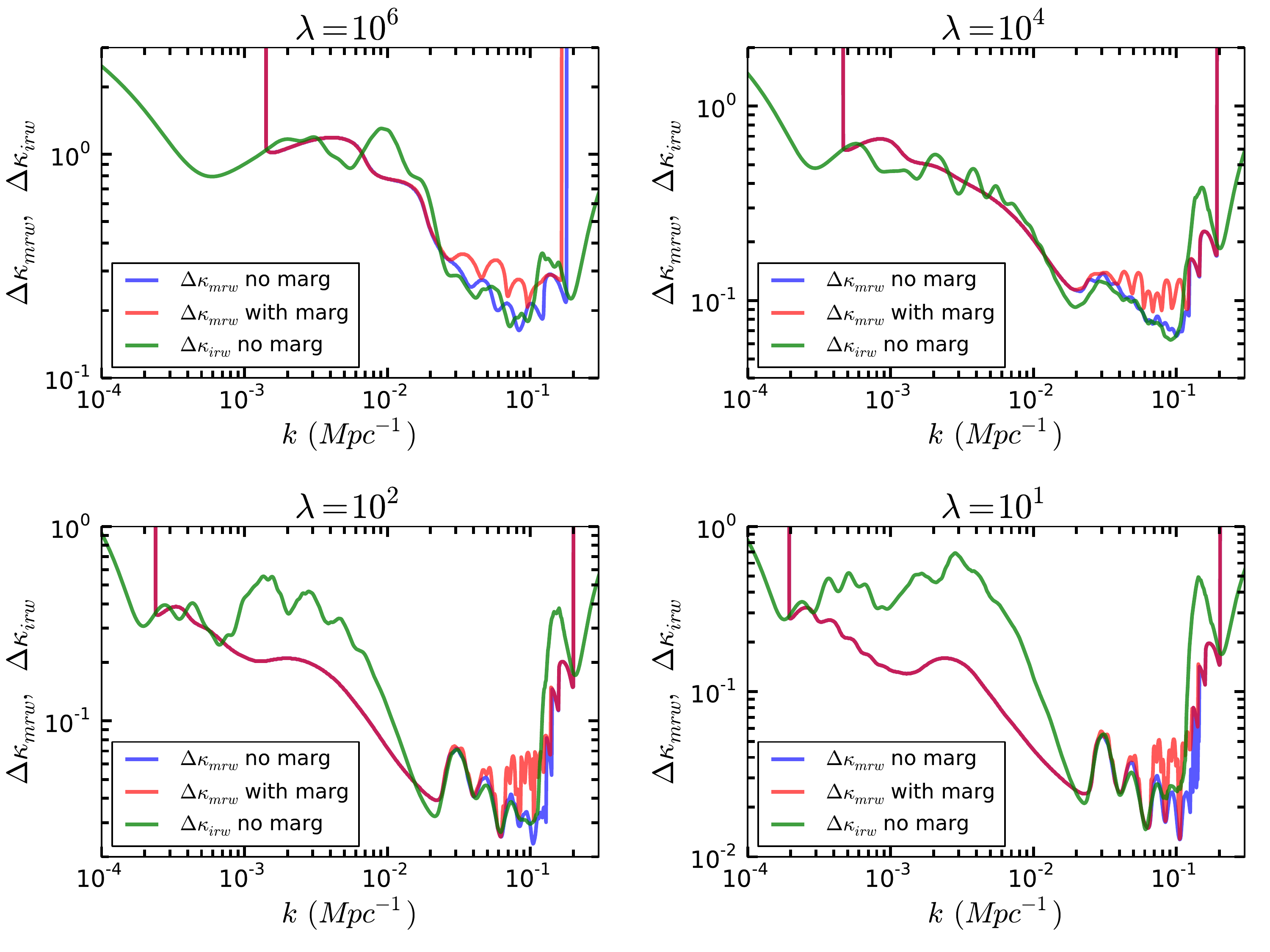}
\caption{
\baselineskip=0pt
{\bf Minimum reconstructible width} as defined in eqn.~(\ref{biasCriterion}) with other parameters fixed (blue) and 
with marginalization including HST and deuterium external data (red). The impulse response width without marginalization has been plotted 
in green for comparison with the minimum reconstructible width.}
\label{bias_length}
\end{center}
\end{figure}

Fig.~\ref{bias_length} 
plots the minimum reconstructible
widths for the marginalized and unmarginalized cases, including the 
impulse response width for the unmarginalized case for comparison.

\clearpage
\section{Reconstruction noise and statistical significance}

The previous section discussed the mean (nonrandom) reconstruction error 
for features in the power spectrum superimposed onto a power
law fiducial spectrum. In this section we examine the random error (having 
zero mean) of the reconstruction resulting from cosmic variance and other 
noise in the determination of the CMB angular power spectrum. We
calculate the noise assuming the null hypothesis that the fiducial
power spectrum is correct. Under this hypothesis, the noise covariance
matrix is given by the expression 
\ba
N_{recon}=
\Bigl(\hat I+R(\lambda ,\alpha )\Bigr)^{-1}
\hat I~
\Bigl(\hat I+R(\lambda ,\alpha )\Bigr)^{-1}
\label{MyError}
\ea
in the presence of the regulator or smoothing penalty. In the absence
of a smoothing regulator, the noise would simply equal $\hat I^{-1}.$ If we had taken
the Bayesian prior in the regulator seriously as a prior over the 
real inputs, the random error would be $\Bigl(\hat I+R(\lambda ,\alpha)\Bigr) ^{-1},$
which is larger than eqn.~(\ref{MyError}).
The difference between the two expressions can be understood 
as a consequence of the smaller than one eigenvalues of the 
operator $\Bigl(\hat I+R(\lambda ,\alpha )\Bigr)^{-1}\hat I.$ 
The corresponding eigenvectors or eigenfunctions correspond
to modes suppressed by the smoothness penalty.
Since we are
primarily interested in assessing the statistical
significance of a potential first detection
against the null hypothesis of a pure power law spectrum,
we adopt the error in eqn.~(\ref{MyError}) in the calculations
below. The operators above include both indices for the 
continuous parameter of $f$ and the other cosmological
parameters $\alpha $ in the case where these are not fixed.
We report on both the random errors
of the cosmological parameters and of the feature
reconstruction. 

\clearpage

\subsection{Random error with other cosmological parameters fixed}

\begin{figure}[!h]
\begin{center}
\includegraphics[scale=.5]{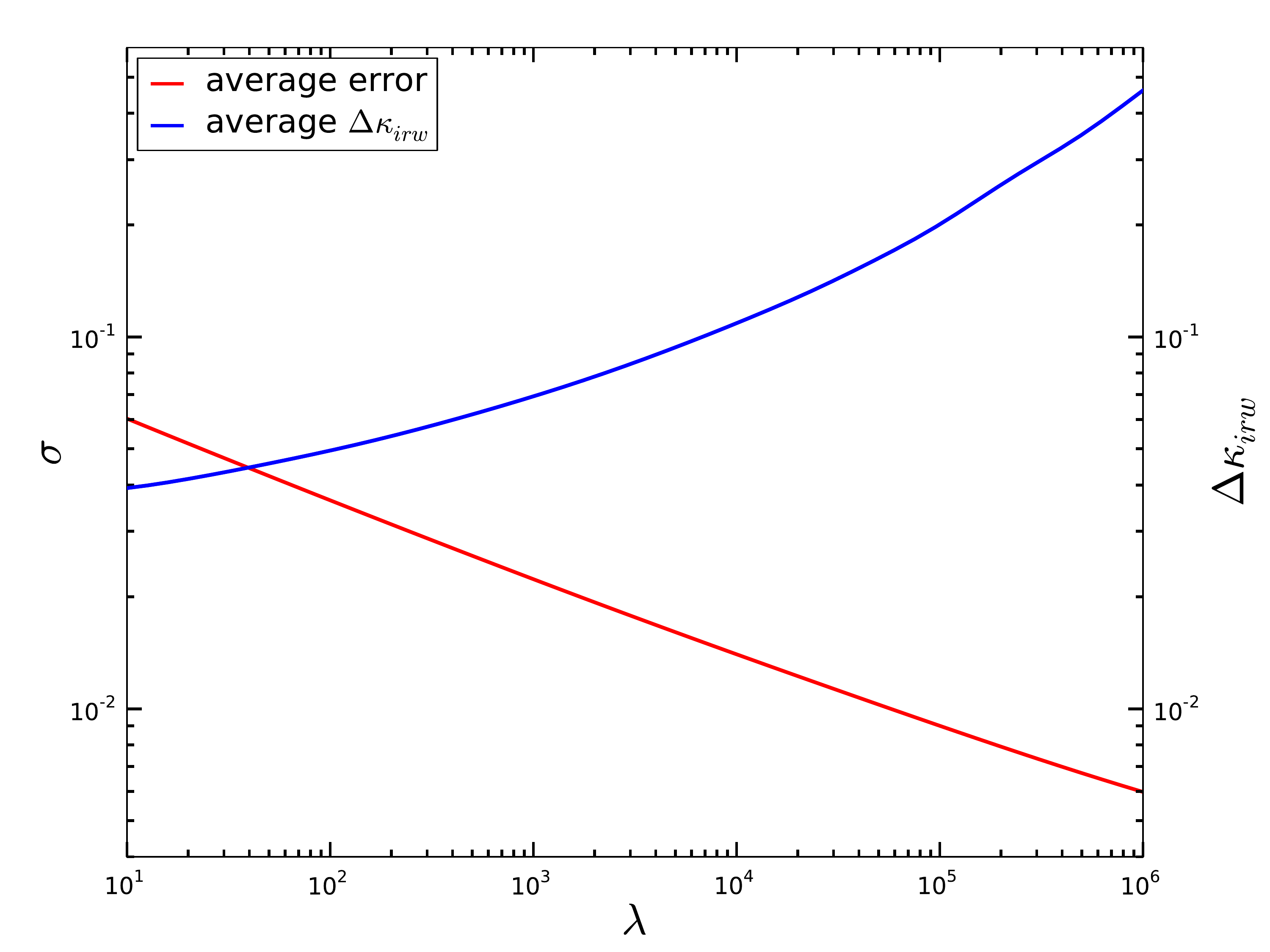}
\caption{
\baselineskip=0pt
{\bf Random error with other cosmological parameters fixed.}
The impulse response width and random error autocorrelation function
averaged over $k \in [0.01,0.1] \, Mpc^{-1}$ is shown as a 
function of the smoothing parameter $\lambda$. 
}
\label{TableOne} 
\end{center}
\end{figure}

\begin{figure}[!h]
\begin{center}
\includegraphics[scale=.45]{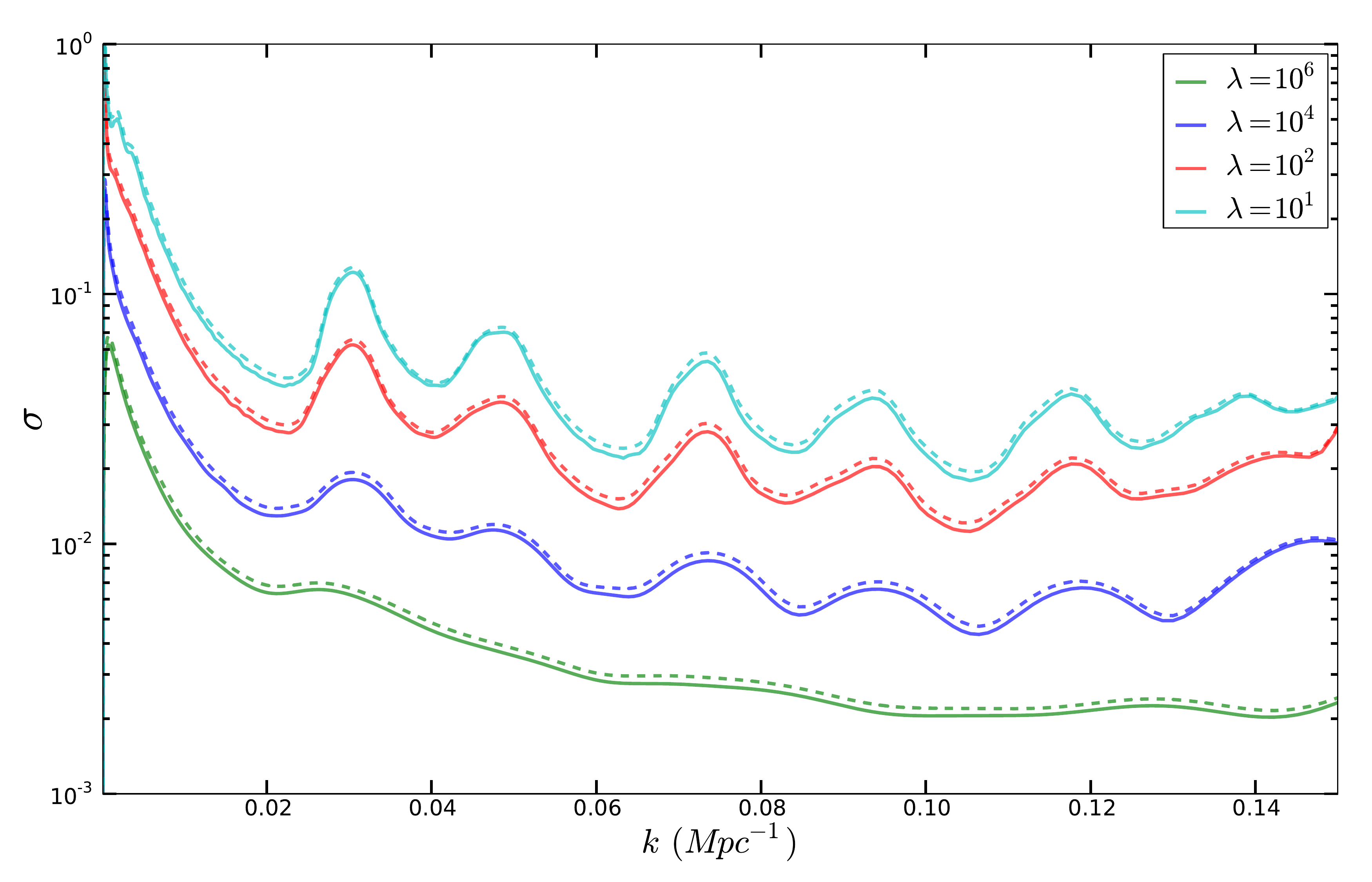}
\caption{
\baselineskip=0pt
{\bf Random error autocorrelation with other parameters fixed.}
The predicted error is dashed while the measured error is solid.
}
\label{muliple_error_unmarg}
\end{center}
\end{figure}

Fig.~\ref{TableOne} shows how the impulse response width and reconstruction 
error (averaged over the $k$ range $[0.01,0.1] \, Mpc^{-1}$) vary as a function of 
$\lambda$. The bias decreases as $\lambda $ is lowered, but this entails a rising
error from reconstruction noise, whose autocorrelation function is given by the
diagonal elements $N_{recon}(k,k)$ as defined in eqn.~(\ref{MyError}). It can be 
seen from the plot that $\sigma \Delta \kappa _{irw}$ is approximately constant.
Fig.~\ref{muliple_error_unmarg} shows the shape of the random error autocorrelation
function. 

The reconstruction method was tested by generating several thousand Gaussian 
realizations of a fiducial $C_{\ell}$ power spectrum calculated using CAMB 
\cite{CAMB} for a given PPS 
and cosmological parameter values.  The features had the Gaussian form 
\begin{gather}
f(k) = A \exp\left[-\frac{(\ln k - \ln k_{c})^{2}}{2 \sigma^{2}}\right]
\end{gather}
parameterized by the width $\sigma$, height $A,$ and center $k_{c}$, chosen 
according to $\lambda$ so that the feature is just wide 
enough to be accurately recovered (as determined by the impulse response width 
in Fig.~\ref{bias_length}) 
and just high enough to obtain a $5\sigma$ deviation from the scale 
invariant spectrum. 
According to Figs.~\ref{bias_length} and \ref{muliple_error_unmarg}, 
the bias and error are minimal around $0.01 Mpc^{-1} < k < 0.1 Mpc^{-1}$, so we 
test the reconstruction with features localized at $k_{c} = 0.05 $ Mpc${}^{-1}$.

\begin{figure}[!h]
\begin{center}
\includegraphics[scale=.45]{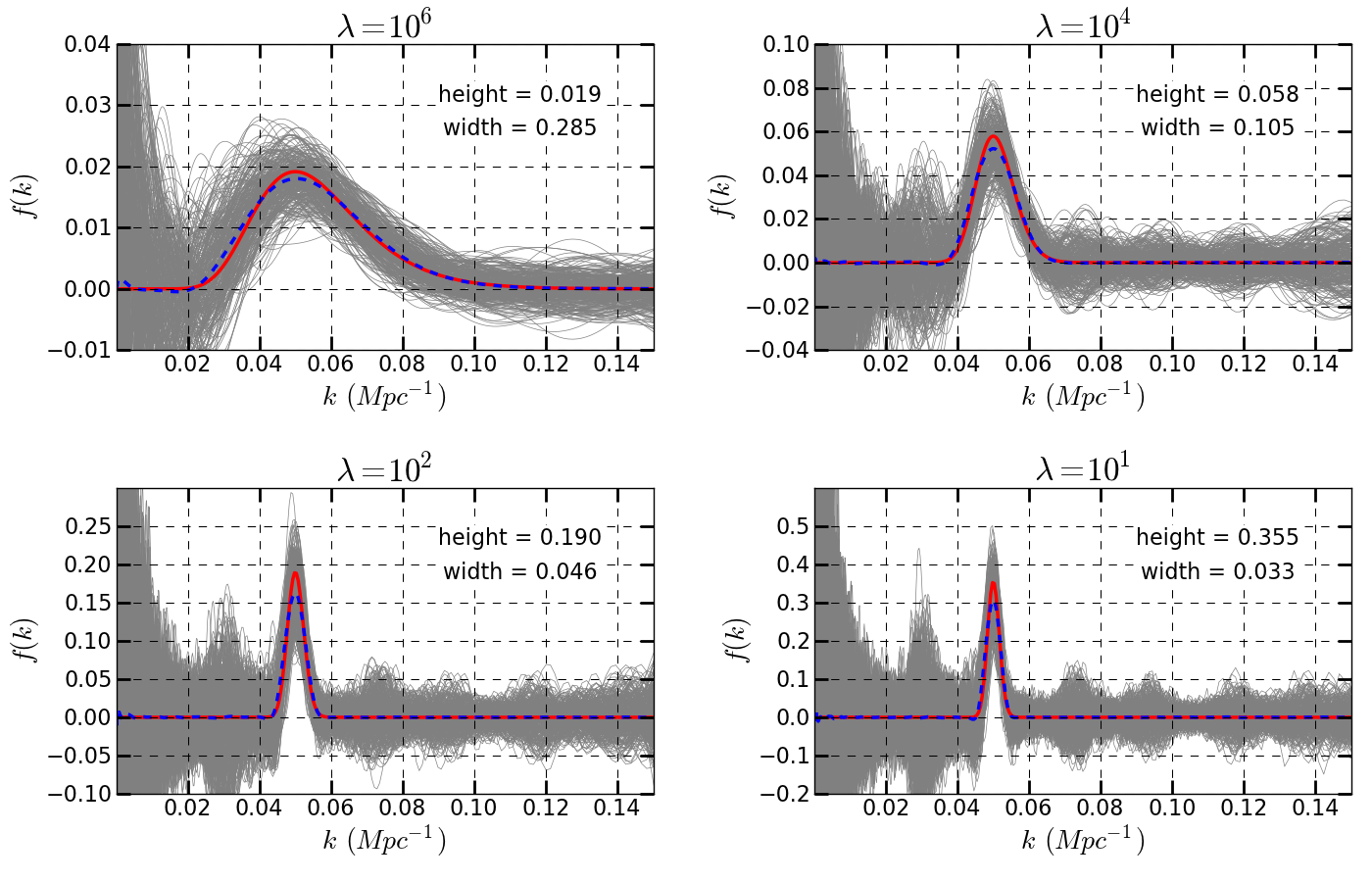}
\caption{
\baselineskip=0pt
{\bf Reconstructed PPS from $10^{4}$ Monte Carlo realizations with other parameters fixed
for four different mock features.} The actual PPS (solid red curve) and 
reconstructions average (dashed blue curve) agree. 
}
\label{muliple_recon_unmarg}
\end{center}
\end{figure}

Fig.~\ref{muliple_recon_unmarg} shows the result of 
$10^{4}$ Monte Carlo simulations of the reconstruction of 4 mock features. 
The expected error in eqn.~(\ref{MyError}) agrees well with the standard 
deviation of the MC reconstructions in Fig.~\ref{muliple_error_unmarg}.

\clearpage

\subsection{Random errors with other cosmological parameters unfixed}

\begin{figure}[!h]
\begin{center}
\includegraphics[scale=.5]{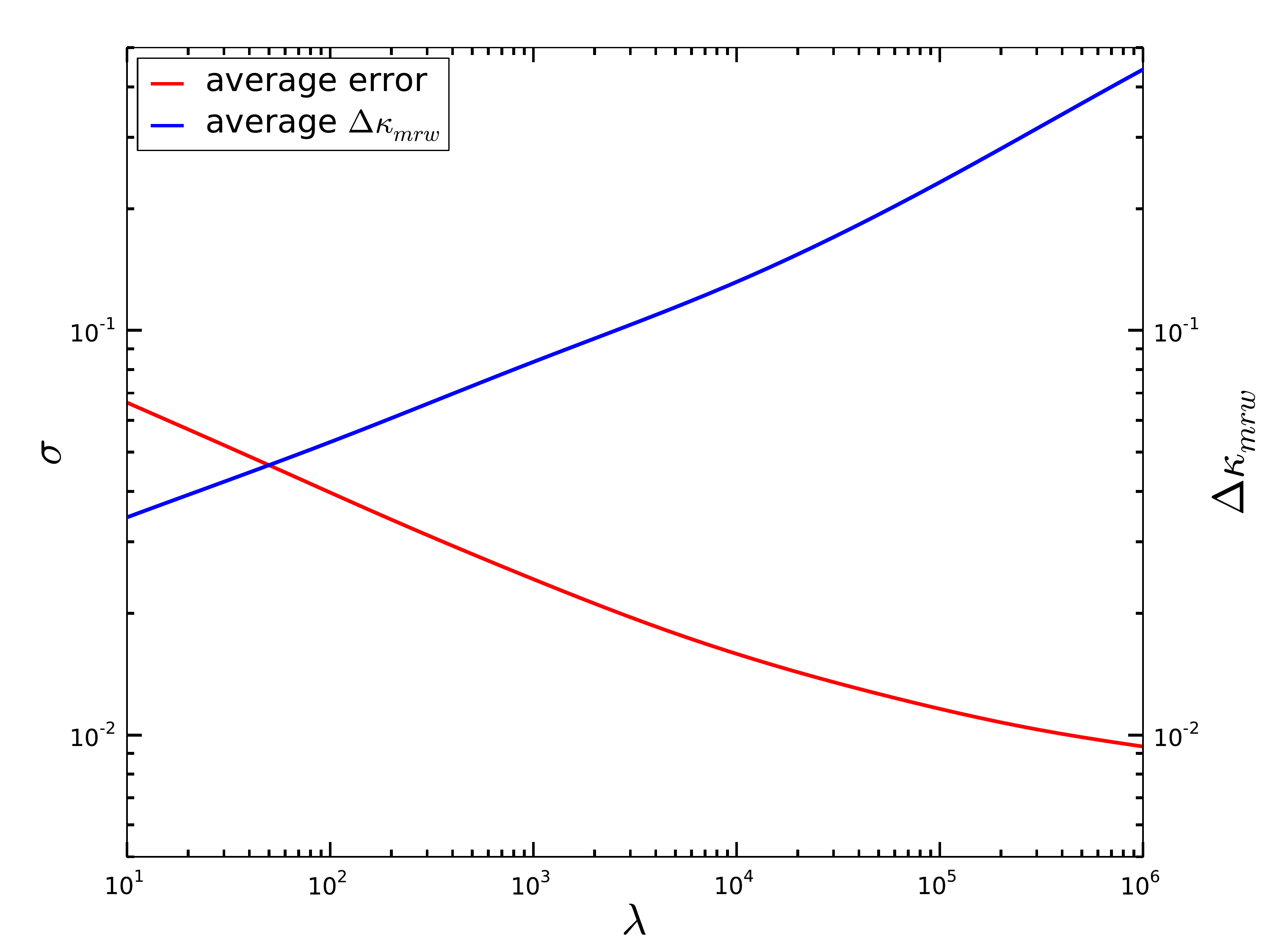}
\caption{
\baselineskip=14pt
\bf
Minimum reconstructible width and random error autocorrelation 
averaged over $k \in [0.01,0.1] \, Mpc^{-1}$ 
with other parameters
uncertain but with external priors on $H_0$ and $\omega _b$
as  
function of $\lambda$.
}
\label{TableTwo} 
\end{center}
\end{figure}

\begin{figure}[!h]
\begin{center}
\includegraphics[scale=0.45]{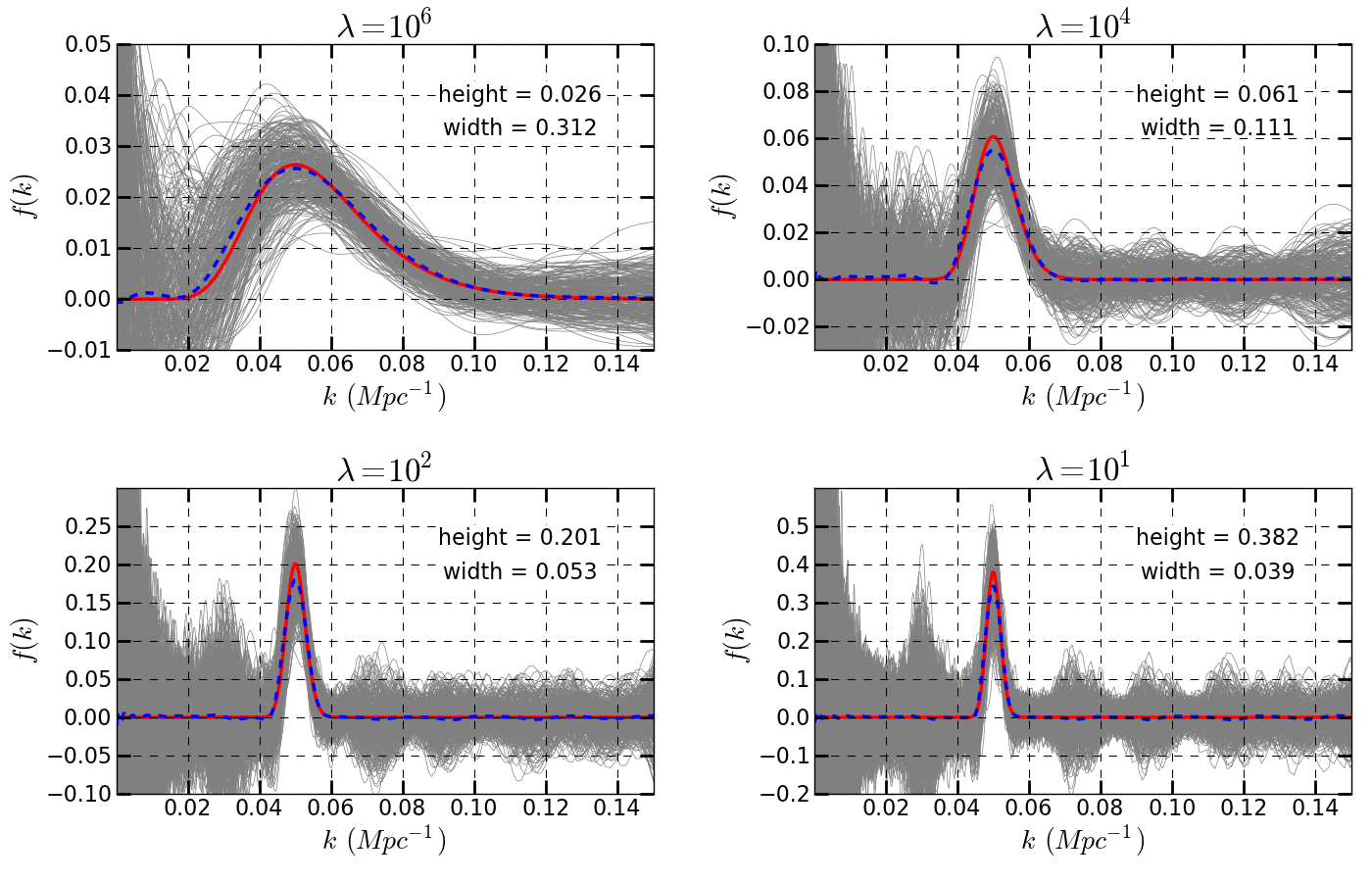}
\caption{
\baselineskip=0pt
{\bf Reconstruction of the PPS from $10^{4}$ Gaussian random realizations 
of the $C_{\ell}$ power spectrum for four different PPS.} The actual PPS is shown as a 
solid red curve while the average of the reconstructions is shown as a dashed blue curve. 
The cosmological parameters have been marginalized over with the HST and deuterium external
data included.}
\label{muliple_recon_marg}
\end{center}
\end{figure}

\begin{figure}[!h]
\begin{center}
\includegraphics[scale=.4]{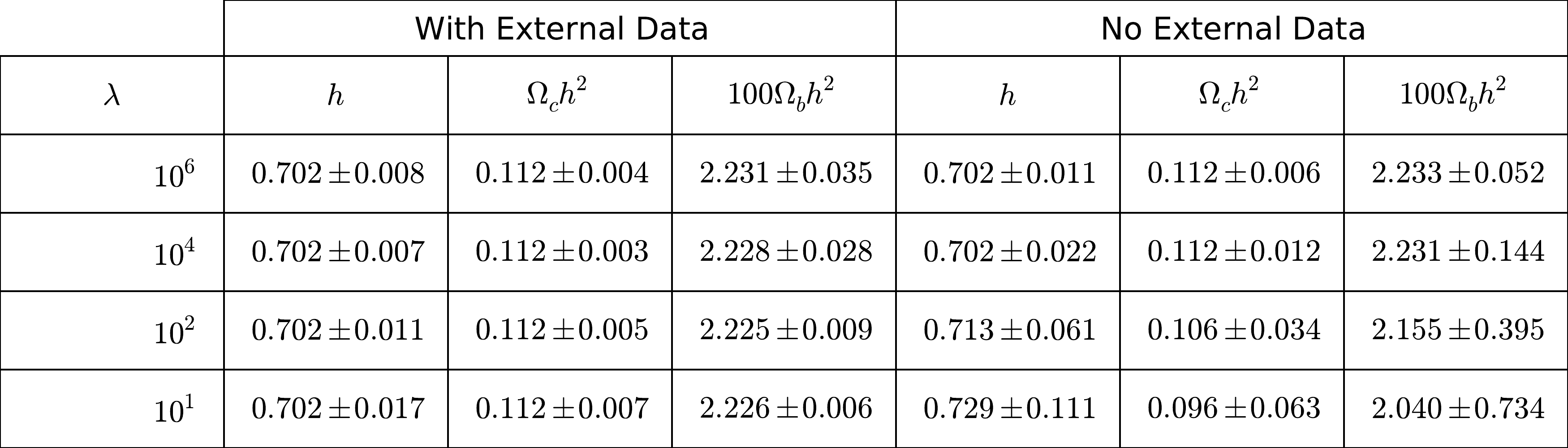}
\caption{
\baselineskip=0pt
{\bf Effect of external priors on $H_0$ and $\omega $ on the cosmological parameters.}
Here for the external data we used the HST and deuterium data error bars but shifted 
the central values to coincide with the WMAP 7-year best fit values. It is the shift 
in these parameters as $\lambda $ is varied that is of interest here.
}
\label{cosmo_param_table}
\end{center}
\end{figure}

In the case with uncertainty in the other cosmological parameters, the reconstruction 
becomes slightly more difficult. 
As in the fixed case, we reconstruct the PPS for four different 
features using $10^{4}$ Gaussian random realizations of a fiducial $C_{\ell}$ power spectrum,
as shown in  
Fig.~\ref{muliple_recon_marg}. 
The shape and position of the feature were chosen as in the 
fixed case except that the minimal reconstructible width is used instead of the impulse response
width for choosing the input feature width. 
Comparing the average of all $10^{4}$ reconstructed spectra 
(dashed blue line) and the actual PPS (solid red line), we see that the PPS is well 
reconstructed. Compared to the fixed case (Fig.~\ref{muliple_recon_unmarg}), 
the errors are slightly larger, particularly for $k$ between 
$0.01$Mpc${}^{-1}$ and $0.1$Mpc${}^{-1}$. The reconstructed cosmological parameters 
with errors are given in Table \ref{cosmo_param_table}, both with and
without external data $H_0$ and $\omega _b$.

Table \ref{cosmo_param_table} shows that without 
external data to constrain $\omega _b$ and $H_0$ (from deuterium abundances and the HST key project
results), the parameters take unreasonable and possibly negative values for small $\lambda .$

\clearpage
\subsection{Graphical representation of reconstruction} 

In this subsection we present a graphical representation of the
reconstructed power spectrum with errors as given in eqn.~(\ref{MyError}).
As we saw, $f(k)$ is not measurable at a point. Rather it is the integral over the kernel 
that is being reconstructed. We therefore indicate error boxes whose width is
equal to the impulse response width for the fixed case
and the minimal reconstructible width for
the case of uncertainty in the other parameters.

\begin{figure}[!h] 
\begin{center} 
\includegraphics[scale=0.45]{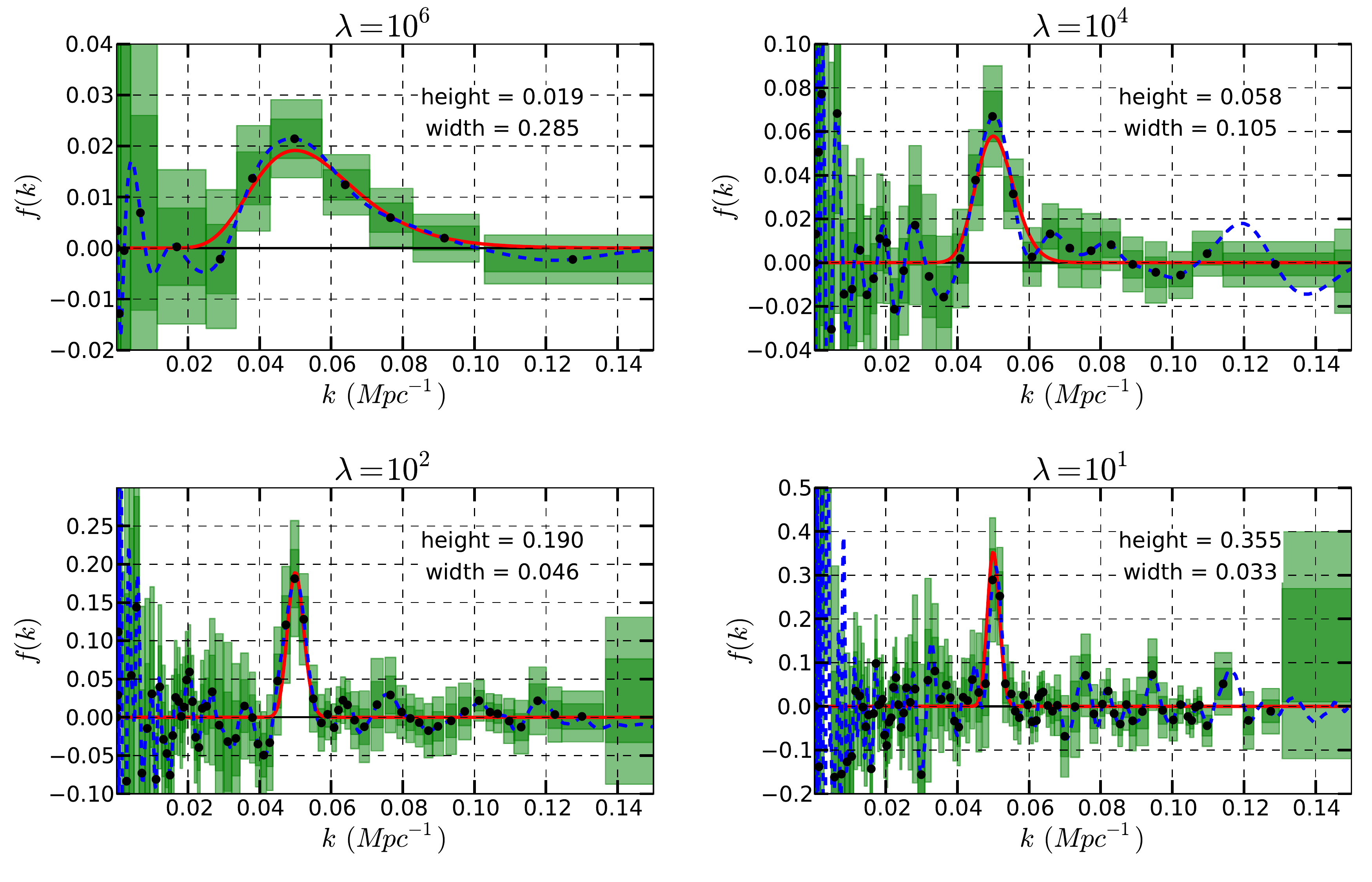} 
\caption{ 
\baselineskip=0pt {\bf Reconstruction of mock features with other cosmological
parameters fixed.} 
The actual feature (solid red) agrees with the average reconstruction 
(dashed blue). Green boxes indicate the error of the 
reconstruction. Their heights represent the $1 \sigma$ and $2 \sigma$ errors
and their widths represent the impulse response width (as evaluated at 
the box center).} 
\label{single_recon_unmarg} 
\end{center} 
\end{figure}

\begin{figure}[!h]
\begin{center}
\includegraphics[scale=0.45]{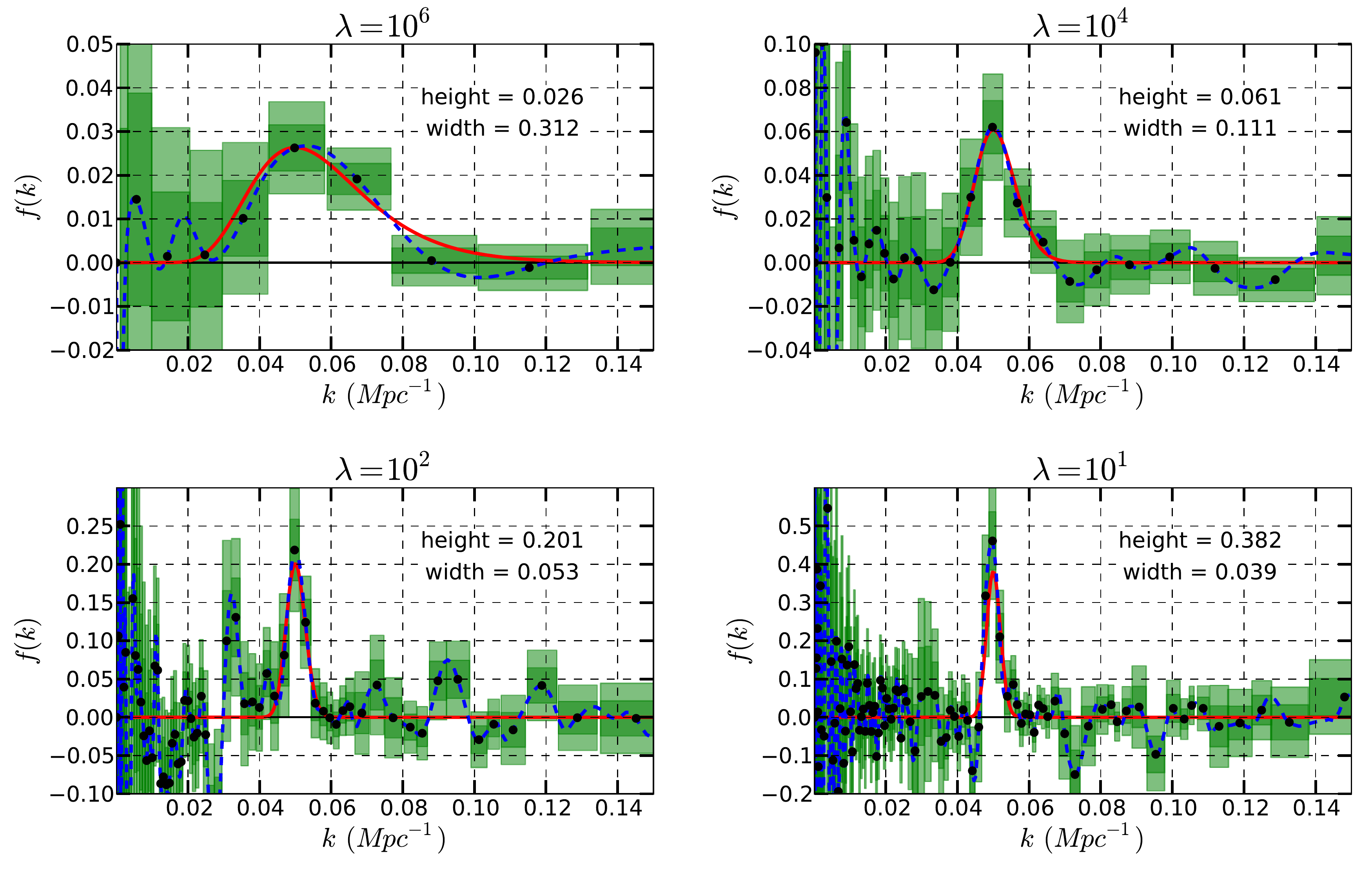}
\caption{
\baselineskip=0pt
{\bf Reconstruction with $h,$
$\omega _b,$ and $\omega _c$ uncertain with HST and deuterium abundance external constraints
included.}
}
\label{single_recon_marg}
\end{center}
\end{figure}

Figs.~\ref{single_recon_unmarg} and \ref{single_recon_marg} illustrates 
the reconstruction of the PPS without the other parameters fixed
and with the other parameters uncertain but with the external data included, 
respectively. The input feature shapes are the same as in
Figs.~\ref{muliple_recon_unmarg} and \ref{muliple_recon_marg}. The $1\sigma$ 
and $2 \sigma$ errors in the reconstruction are represented by the height of the green 
shaded boxes. Thus the PPS has been successfully reconstructed if the 
actual PPS passes through most of the error boxes, which is the case
in Figs.~\ref{single_recon_unmarg} and \ref{single_recon_marg}. 

\clearpage
\subsection{Non-Gaussian corrections}

The preceding discussion assumed a
quadratic likelihood. However the actual likelihood
is nonlinear mainly because of non-linearity
in the low-$\ell $ likelihood for the
$C_\ell $'s 
and
the nonlinear dependence of the predicted
$C_\ell $'s 
as a function of the cosmological
parameters. Once the cosmological parameters
have been fixed, the derivative of the likelihood
with respect to variations in the power
spectrum is easy to compute
analytically. Moreover, many of the
high-$\ell $ likelihoods for a theoretical 
$C_\ell $ spectrum 
taking into account partial sky coverage,
nonuniform instrument noise, and removal
of high-$\ell $ residual galactic and extragalactic
contamination are based on approximations
for which derivatives of the likelihood
with respect to changes in the $C_\ell $'s may be
calculated with simple code modifications.
For feature reconstruction it is the high-$\ell $
likelihood, and not the low-$\ell $ likelihood,
where a pixel-based approach is required, that
is most relevant. Consequently, for obtaining the
maximum likelihood solution, it is feasible
and simple to apply the Newton-Raphson
root finding algorithm to obtain the exact
ML reconstruction within a few iterations despite
the massive increase in the number of parameters.
Accounting for non-Gaussianity in the fluctuations
about the ML reconstruction is more difficult. MCMC
methods are not feasible because there are too
many parameters. There is likely to be mild
non-Gaussianity impacting on the analysis in
Section 4. This is an issue currently under
investigation.

\clearpage
\section{Conclusion}

In this paper we have explored a model-independent scheme 
for reconstructing possible features in the primordial
power spectrum based on a penalized likelihood. After
exploring the structure of the available information, 
we study the properties of the reconstruction for
a range of degrees of smoothing. What features in
the primordial power spectrum
can be reconstructed and with what statistical significance
is quantified for a collection of representative shapes.
We found that very narrow features cannot be
reliably reconstructed due to confusion with
the cosmological parameters even when 
external data is used to fix $H_0$ and $\omega _b.$
It will be interesting to see whether future data
will uncover significant deviations from a power
law adiabatic primordial power spectrum.

\clearpage


{\bf Acknowledgements:} The authors acknowledge the use of CAMB
for calculating the transfer functions used in the calculations
in this paper. (See http://www.camb.info/) CG was supported by
a Centre National des \'Etudes Spatiales (CNES) postdoctoral fellowship. 

\clearpage

\bibliography{PPS-paper}


\nocite{Hannestad:2003zs}
\nocite{Bridle:2003sa}


\nocite{Verde:2008zza}
\nocite{Peiris:2009wp}
\nocite{Sealfon:2005em}


\nocite{Mukherjee:2003yx}
\nocite{Mukherjee:2003ag}
\nocite{Mukherjee:2005dc}


\nocite{Hu:2003vp}
\nocite{Leach:2005av}


\nocite{Shafieloo:2007tk}
\nocite{Nagata:2008tk}
\nocite{Nagata:2008zj}
\nocite{Hamann:2009bz}
\nocite{Kogo:2003yb}
\nocite{Shafieloo:2003gf}
\nocite{TocchiniValentini:2004ht}
\nocite{TocchiniValentini:2005ja}
\nocite{Ichiki:2009zz}

\end{document}